# Fundamental Investigation of Reactive-Convective Transport: Implications for Long-Term Carbon dioxide (CO$_2$) Sequestration

Md Fahim Shahriar[1], Aaditya Khanal[1*]

[1]The Jasper Department of Chemical Engineering, The University of Texas at Tyler

Corresponding author E-mail address: aadityakhanal@uttyler.edu

**Abstract**

The density-driven convection coupled with chemical reaction is the preferred mechanism for permanently storing CO$_2$ in saline aquifers during its injection into deep saline aquifers. The flow behavior and reaction mechanism of the injected CO$_2$ depends on the geochemical profile of the formation rock and brine. This study uses a 2D visual Hele-Shaw cell to evaluate and visualize the density-driven convection formed due to gravitational instabilities, also known as Rayleigh-Taylor instability. The primary goal of the experiments is to understand the various mechanisms for the mass transfer of gaseous CO$_2$ into brine with different initial ionic concentrations and flow permeability. Moreover, the impact of CO$_2$ flow rates, injection locations, reservoir dipping angle, and permeability heterogeneity is also investigated. We observed that the presence of salts resulted in earlier onset of convection and a larger convective finger wavelength than the case with no dissolved salts. In addition, experimental data showed a higher lateral mixing between CO$_2$ fingers when dipping is involved. The visual investigation also revealed that the CO$_2$ dissolution rate, measured by the rate of the convective fingers advance, depends on the type and concentration of the ions present in the brine. The CO$_2$ dissolution for solutions with varying salt dissolved, indicated by the area of the pH-depressed region, is observed to be 0.38-0.77 times compared to when no salt is present. Although convective flow is slowed down in the presence of salts, the diffusive flux is enhanced, as observed from both qualitative and quantitative results. Moreover, the reduced formation permeability, introduced by using a flow barrier, resulted in numerous regions not being swept by the dissolved CO$_2$, indicating an inefficient dissolution. We also investigated the effect of discrete high conductivity fractures within the flow barriers, which showed an uneven vertical sweep and enhanced flow channeling. Lastly, the parameters regarding CO$_2$ leakage risk during storage are identified and discussed. Moreover, the quantitative comparison of the test cases obtained from image analysis elucidated the convection dynamics of the CO$_2$ storage process. The fundamental insights from this study are applicable for optimizing and improving the geo-sequestration of CO$_2$ in subsurface formations saturated with brine.

# 1. Introduction

The atmospheric concentration of carbon dioxide ($CO_2$) in May 2021 was recorded as 419.5 parts per million (ppm), which is approximately 50% higher than at the beginning of the industrial revolution [1]. The growing concerns about $CO_2$ emissions have led to investigations of possible carbon capture and storage (CCS) methods. $CO_2$ geological storage in depleted oil reservoirs or saline aquifers is one of the preferred CCS methods to capture emissions from large point sources [2]. $CO_2$ geological storage has four primary $CO_2$ trapping mechanisms: structural, residual, dissolution, and mineral trapping, as shown in Fig. 1 [1]. Capillary or residual trapping rate is the highest at the beginning of the $CO_2$ storage period; however, dissolution trapping becomes more dominant throughout time (**Fig. 1**). Dissolution trapping captures almost two-thirds of $CO_2$ injected in the storage volume [3–5]. Even though molecular diffusion of $CO_2$ in brine is slow, the rate of dissolution trapping is accelerated by other mechanisms, including density-driven convection or Rayleigh-Taylor instability, dispersion, and advection [6,7].

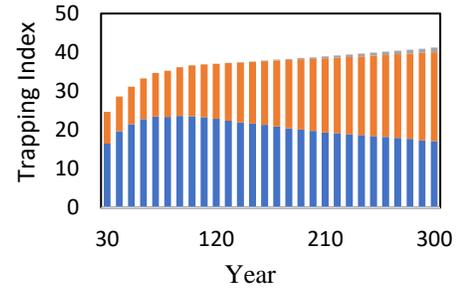

**Figure 1.** Fraction of $CO_2$ sequestered by residual trapping (blue), dissolution (orange), and mineralization (gray) (Modified from Khanal and Shahriar [1]). Dissolution is the primary storage mechanism especially with the passage of time.

Several factors can affect the convective-dissolution phenomenon for different geological sites and, therefore, needs to be considered for accurate $CO_2$ storage prediction. For example, density-driven convection is reduced for geological storages with lower vertical permeability, resulting in advection becoming the dominant force with increased transverse mixing [3,8]. Nevertheless, the effect of density-driven convection in $CO_2$ geological storage is significant, as it prevents $CO_2$ leakage by sinking $CO_2$ rather than rising to shallower formations [3]. Furthermore, since $CO_2$ dissolution into brine increases the density of brine on the order of 0.1-1%, the density gradient between the top layer (brine-$CO_2$ solution) and the layer immediately below (pure brine) induces density-driven natural convection under gravitational instabilities [1–3].

Various studies have identified and extensively studied different aspects of the transport mechanisms during the density-driven convection or Rayleigh-Taylor instability, including diffusion period, convective period, and constant flux [9–19]. Hele-Shaw cell, a simple structure usually formed by a narrow gap between two transparent flat plates, has been used in multiple experimental setups to visualize and mimic the Rayleigh-Taylor instability or Rayleigh convection (RC) formed during $CO_2$ storage in different geological structures [12,20–23]. Kneafsey and Pruess [24] conducted laboratory visualization studies and quantitative $CO_2$ absorption tests in transparent Hele-Shaw cells to investigate the dissolution-induced density-driven convection phenomenon. The quantitative measurements showed that the density-driven convection initiated faster than predicted. Backhaus et al. [25] studied the density-driven convection for a lighter fluid (water) placed over a heavier fluid (propylene glycol). The initial instability and quasi-steady-state were explained by analyzing the convective time and velocity scales along with the finger width and the rate of mass transport. The test was conducted at standard atmospheric conditions with a Rayleigh number (Ra) of 6000-90,000 [25]. Meanwhile, a smaller Ra range (100 < Ra < 1700) was adopted in the works of Slim et al.

[26]. Potassium permanganate (KMnO$_4$) in water was used as an analog for CO$_2$ in brine at atmospheric conditions, describing the dissolution-driven convective behavior from the first contact up to 65% average saturation.

Developing scaling relationships, correlations and models can provide important insight into convection dissolution properties of CO$_2$ for different geological storage [2,27–30]. Motjaba et al. [27] developed two scaling relationships, one between Rayleigh and Sherwood numbers and the other between Rayleigh numbers and CO$_2$ convective flux. Moreover, density-driven convection behavior was observed using visualization techniques and quantitative experiments (under 3.45 MPa and 182 < Ra <20860). Robust scaling relations between compensated flux and transition times between successive regimes in the system for different salt types (NaCl and CaCl$_2$) were examined by Mahmoodpour et al. [31]. The results showed that different salt types affect both the short and long-term dynamics of convective dissolution. Faisal et al. [28] obtained correlations between the Rayleigh number and the mass of total dissolved CO$_2$. The mass of dissolved CO$_2$ was determined under atmospheric conditions (with 3277.88 < Ra < 36,420.87) using a catalytic combustion-based total carbon analyzer (TC-analyzer). Other useful scaling laws, including the onset of the convection, and wavelength of the initial convective instabilities, were also identified and discussed [4,32,33]. Tani et al. [34] analyzed the one-phase problem on radial viscous fingering in a Hele-Shaw cell. The mathematical analysis was performed by modifying Stefan's problem and justifying the time-derivative's vanishing coefficient in the parabolic equation.

Different visualization techniques, including the Schlieren method, particle image velocimetry (PIV), laser-induced fluorescence (LIF), and interferometry method, have been adopted to observe the formation and growth of convective finger structures in Rayleigh convection [2,3,35,36]. Zhang et al. [2] presented a vortex model of CO$_2$ adsorption into the water to characterize the interfacial mass transfer coefficient for the continuous convective period. The study adopted particle image velocimetry (PIV) and laser-induced fluorescence (LIF) to calculate the solute concentration distribution and instantaneous liquid velocity in the Hele-Shaw cell. In addition, this study considered the 3D simulation approach, which provided a critical overview of how the regions inside and outside the convective fingers have enhanced interfacial mass transfer by reducing the thickness of concentration boundary layers. Moreover, another recent work by Zhang et al. [37] used a different experimental setup using the UV-induced fluorescence method to investigate gas-liquid interphase mass transfer in a Hele-Shaw cell. This experimental approach was low cost and more sensitive to changes than laser devices for pH-sensitive fluorescers.

Mahmoodpour et al. [31] provided critical insight into visualizing the dissolution-driven convection at high-pressure conditions (up to 535.3 psi). They devised a novel Hele-Shaw apparatus withstanding high pressure and presented CO$_2$ dissolution-driven convective behavior in a confined brine-saturated porous medium. Tang et al. [3] designed a Hele-Shaw cell rated to 70 MPa and Ra of 346 to investigate the convection parameters, including critical onset time of convection, dissolution rate, and gravitational instabilities. This study used the micro-schlieren technique to conduct the visual inspection. Pressure-volume-temperature (PVT) testing was conducted at 293.15 to 423.15 K and pressure ranging from 14 to 24 MPa.

The chemical composition of the CO$_2$ storage site also significantly affects the reactive transport of dissolved CO$_2$, as observed in the works of Thomas et al. [38]. Their work investigated the dissolution of CO$_2$ into an aqueous solution of bases MOH, where $M^+$ is an alkali metal cation. For bases MOH, the convection is enhanced for counter-ion $M^+$ sequence of $Li^+ < Na^+ < K^+ < Cs^+$. The experimental investigation revealed that the concentration of base in solution strongly impacts the nonlinear finger instability, where higher concentration leads to faster instability and shorter time for onset of convection. Furthermore, despite $M^+$ ions not actively participating in the geochemical reactions during the dissolution process, the nature of different $M^+$ ions vary in the instability development.

Loodts et al. [39] observed the effect of pressure, temperature, and NaCl concentration on $CO_2$ dissolution properties. Their study suggested that increasing $CO_2$ pressure or reducing temperature or salt concentration leads to higher convective instability. However, temperature has a minimal effect on $CO_2$ dissolution properties, so controlling the temperature is not essential for the reproducibility of experimental studies [39]. Thomas et al. investigated the effect of salinity by the dissolution of gaseous $CO_2$ in pure water, Antarctic water, and 0.5-5 M NaCl dissolved in water [40]. The results showed that higher salt concentration delays the formation of instabilities, resulting in delayed onset of convection. Moreover, increased convection pattern wavelength and decreased fingers' velocity and the growth rate increased the salt concentration. Kim and Kim [41] derived and solved linear stability equations for the effect of chemical reactions in an initially quiescent vertical Hele-Shaw cell. Their nonlinear numerical simulation showed that chemical reactions enhance the diffusive flux; however, by retarding the onset of buoyancy-driven convective motion, convective flux is weakened.

Formation dip angle is another key factor of consideration for safely storing $CO_2$ on subsurface geological sites, as it significantly impacts spatial migration distribution during $CO_2$ dissolution [42–44]. For larger dip angles, the supercritical $CO_2$ phase could change to a gas phase during upward migration in the reservoir up-dip direction, where the reservoir formation temperature and hydrostatic pressure are lower [43]. As a result, reservoirs with higher dip angles have more chance of $CO_2$ leakage during geological storage. Jang et al. [42] simulated the effect of dip angle and salinity of $CO_2$ storage. For formation dip angles of 0°, 5°, and 10°, the migrated $CO_2$ distances were 60%, 73.3%, and 86.7%, respectively, compared to a 15° dip angle in the 200$^{th}$ year of $CO_2$ migration. Therefore, with a larger formation dip angle, there is a higher possibility of spatial $CO_2$ migration. They concluded that reservoirs with higher dip angle and salinity have low $CO_2$ geological storage safety. Wang et al. [43] observed similar effects of formation dip, where the total $CO_2$ storage amount is inversely proportional to the formation dip angle. The impact of dip angle is more prominent in storage reservoirs with higher porosity and permeability [43]. As Jing et al. [35] observed, higher salinity and high dip angle are not conducive to $CO_2$ geological storage. However, the effect of salinity is observed to be more significant than that of dip angle on the $CO_2$ liquid phase mass fraction.

During the injection of $CO_2$ in deep saline aquifers, the natural fractures present in the formation may propagate, or new fractures may be induced in the reservoir. The fracture networks in a hydrocarbon reservoir play a vital role in fluid transport from the pores to the wellbore as they are significantly more conductive than the matrix [45–48]. The same principle is applicable during the $CO_2$ sequestration operation, which makes it difficult to predict the movement of plumes during the injection of $CO_2$ in fractured porous media. Hence, natural and induced fracture networks in the geological storage sites should also be considered to predict $CO_2$ subsurface movement. Due to the opening and closing of the fractures, the reservoir properties also deviate from the value measured from the core analysis. The highly permeable conductive fracture networks can act as a pathway for fluid movement, potentially allowing $CO_2$ migration to neighboring aquifers or the surface through the cap rock [49,50]. Bond et al. [49] demonstrated improved $CO_2$ migration prediction by incorporating structural geological fractures in the model. Knowing the spatial distribution of fractures, their orientation, conductivity, and overall contribution to the effective permeability are desirable for geological sites whose permeability is controlled mainly by faults and fractures [50]. March et al. [51] presented a model for $CO_2$ storage in naturally fractured reservoirs, showing the importance of selecting an appropriate injection rate to prevent "early spill": the fast flow of $CO_2$ through the highly permeable fracture storage before any significant $CO_2$ storage.

Despite the considerable investigative analyses on $CO_2$ dissolution-driven convection using the Hele-Shaw cell, there is still a lack of studies that visualizes $CO_2$ dissolution with a dipping angle involved. Moreover, the effect of varying $CO_2$ flow rates and injection points needs further insight in terms of quantitative analysis. Furthermore, despite the considerable theoretical model and experimental investigation on the effect of salinity in $CO_2$ dissolution, most of the work considers NaCl and ignores the presence of other salts. Lastly, to the best of our knowledge, the effect of fractures in heterogeneous media is yet to be presented. This experimental study critically investigates the effect of these factors to address this knowledge gap. Furthermore, this study discusses how the findings of this work can be translated into improving storage efficiency by providing key insight into $CO_2$ convection dynamics.

The remainder of the paper is organized as follows. **Section 2** provides the experimental methods, image processing sequence, and a brief overview of the experiments considered. **Section 3** presents the qualitative visualization of $CO_2$ dissolution in different homogeneous and heterogeneous media. Additionally, the effect of fractures on $CO_2$ dissolution is also discussed. **Section 4** presents our findings in terms of quantitative data. Lastly, **Section 5** is devoted to the discussion, and **Section 6** presents the main conclusion of this study.

## 2. Experimental Method

### 2.1. Experimental Setup and Materials

Experiments are performed in a Hele-Shaw cell composed of two transparent 0.5 in (12.9 mm) thick plexiglass separated by precision silicone shims with a thickness of 1 mm along the sidewalls. The front plexiglass panel was drilled at the bottom of the cell and was fitted with a ball valve with a diameter of 0.25 in (6.35 mm). The internal cell dimensions were a length and height of 259 mm and 284 mm, respectively. The height (H) of the water column is 243 mm, as shown in **Fig. 2a.** This port was used to fill and drain the reactor of the experimental fluid. At the top, three holes at the center of the front plexiglass were drilled so that the $CO_2$ could be securely injected into the cell using an 18-gauge dispensing needle, as shown in **Fig. 2a**. A digital camera (Nikon D7000 with 50 mm lens) was focused on the cell to take pictures at 20 seconds intervals for 2.5 hours. We maintain a 4 cm distance from the injection point to the top of the water to avoid disturbing and introducing shear stress at the interface. The experiments were conducted using the same protocol to ensure a similar controlled environment. The schematic of the experimental setup is presented in **Fig. 2b.**

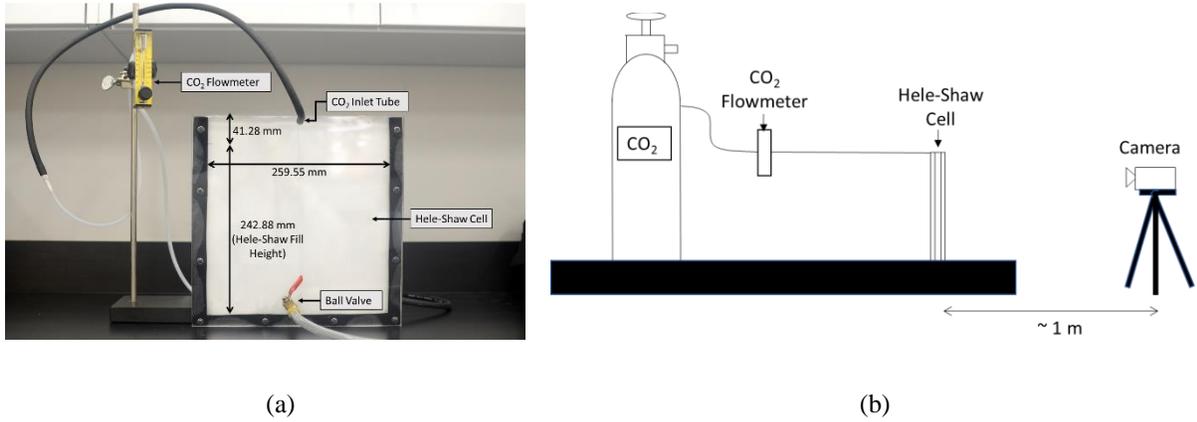

**Figure 2.** (a) The Hele-Shaw cell with dimensions and (b) Schematic of the experimental setup

The flow of incompressible fluid in a Hele-Shaw is governed by the Navier-Stokes equation, coupled with the species conservation equation for cases with chemical reactions [28,52]. The conservation equations for mass, momentum, and species also implicitly consider the Boussinesq approximation, which assumes that the solute concentration affects the local density without causing an expansion or contraction of the fluid [53]. When the aperture between the parallel plates is small compared to height, the flow environment is mathematically analogous to the Darcy flow in isotropic porous media. The permeability of the Hele-Shaw cell, $k$, with a cell aperture of $b$ (for this study, 1 mm), can be obtained from the fundamental conservation laws as [54]:

$$k = \frac{b^2}{12} = \frac{0.001^2}{12} = 8.33 \times 10^{-8} \text{ m}^2 \quad (1)$$

The corresponding Rayleigh number (Ra), a dimensionless number characterizing the system by expressing the ratio of free convection to diffusion, is calculated using the permeability ($k$) of the porous medium and is represented as shown in **Eq. 2**:

$$\text{Ra} = \frac{\Delta \rho g k H}{\mu \varphi \, D_{CO_2}} \quad (2)$$

Where $g$ is the acceleration of gravity [m/s$^2$], $H$ is the height of the water column [m], $\mu$ is the dynamic viscosity of water [ kg/m.s], $D_{CO_2}$ is the molecular diffusion coefficient of $CO_2$ in water [m$^2$/s], $\Delta \rho$ is the increase in density due to $CO_2$ dissolution [kg/m$^3$], $k$ is the permeability of the medium [m$^2$], and $\varphi$ is porosity. For our experimental cases without any heterogeneity, $\varphi = 1$. For the heterogeneous cases, porosity, permeability, and corresponding Rayleigh number are calculated and presented in Section 3.2. We obtain a value of Ra = 41782 for the homogeneous base cases conducted in this study. It should be noted that the Rayleigh number will decrease slightly with an increase in salinity, as observed by Thomas et al. [40]. In their experiment, for salinities of 1M and 2M NaCl, the Rayleigh number decreased by 0.77% and 1.5%, respectively, compared to the Rayleigh number of the water system. We expect salinity to cause a similar decreasing pattern in Rayleigh number for cases with dissolved salts. The values for calculating the Rayleigh number were adopted from Faisal et al. [28], who used a bromocresol green pH indicator at a room temperature of 22 °C, similar to our experiments. As other theoretical studies and experimental investigations observed, for Ra ≥ $4\pi^2$, natural convection is the predominant flow [28,55]. All the experiments in this study meet that criterion for prevailing convective flow.

## 2.2. Experimental Procedure and Image Processing

The parameters considered for image analysis are the number of fingers formed, average finger length (mm), average wavelength (mm), area of pH depressed region ($mm^2$), and standard deviation of finger length (mm). Finger length is calculated as the vertical distance from the interface to the tip of the finger. In contrast, the average wavelength is the interface length divided by the number of fingers.

We followed the same image processing sequence adopted in the work of Kneafsey et al. [20]. Using ImageJ FIJI, we subtracted the initial condition (no $CO_2$) applied from all the images, resulting in images with less background noise in RGB format. Next, the RGB images were split into three color channels: red, green, and blue. The blue color channel containing mostly noise was discarded, while the two other channels were added, as shown in **Fig. 3**. Lastly, we inverted the grayscale lookup table changing dark to light, making the $CO_2$-dissolved pH-depressed regions darker. Additional image processing and noise removal were performed on some images for better clarity.

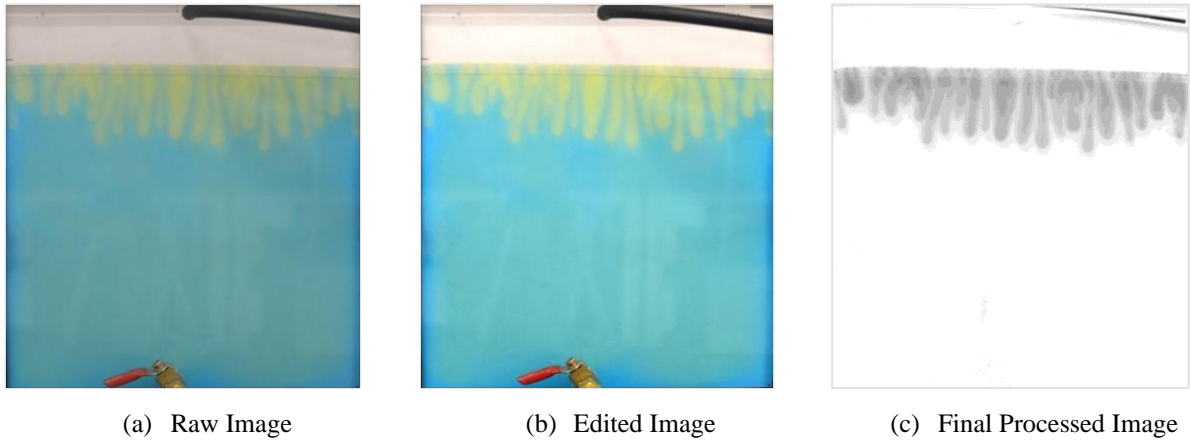

(a) Raw Image   (b) Edited Image   (c) Final Processed Image

**Figure 3.** Image Processing sequence showing (a) unedited image from the camera, (b) edited image with enhanced contrast, (c) grayscale image, which shows the final image used for analysis.

Before performing the experiments (**Table 1**), we conducted a repeatability test by comparing the qualitative and quantitative results of three runs for a test case where $CO_2$ was injected at a steady flow rate of 0.59 L/min into the middle port of the Hele-Shaw cell (**Fig. 2a**) filled with a solution prepared by mixing deionized, de-aired water with 0.0114% w/v Bromocresol Green (BCG). This solution is referred to as the *Control Fluid* (CF) in this study. BCG is a pH indicator that changes color from blue at a pH above 5.4 to yellow at a pH of around 3.9 [20]. The application of pH indicators is a widely accepted method for studying convective mixing [20,21,28,56]. Thomas et al. [57] indicated that the color indicator does not significantly affect the development of convective dynamics. Additionally, Taheri et al. [58] reported no change in water's properties for a small amount of BCG. The solution in their experiment was prepared by mixing distilled deionized water with 0.025 wt. % BCG. Therefore color-based pH indicators can be used, with caution, when trying to quantify convective-dissolution properties [57,58]. The Hele-Shaw cell was filled at approximately 85% with 63 mL of CF.

The qualitative results showing the convective pattern for the same time scale are presented in **Fig. 4,** which show good repeatability for the finger size, number, and overall patterns when identical experimental conditions are used for the runs. We also calculated the 95% confidence interval for various parameters that characterize the fluid flow presented later in the quantitative measurement section (**Section 4)**.

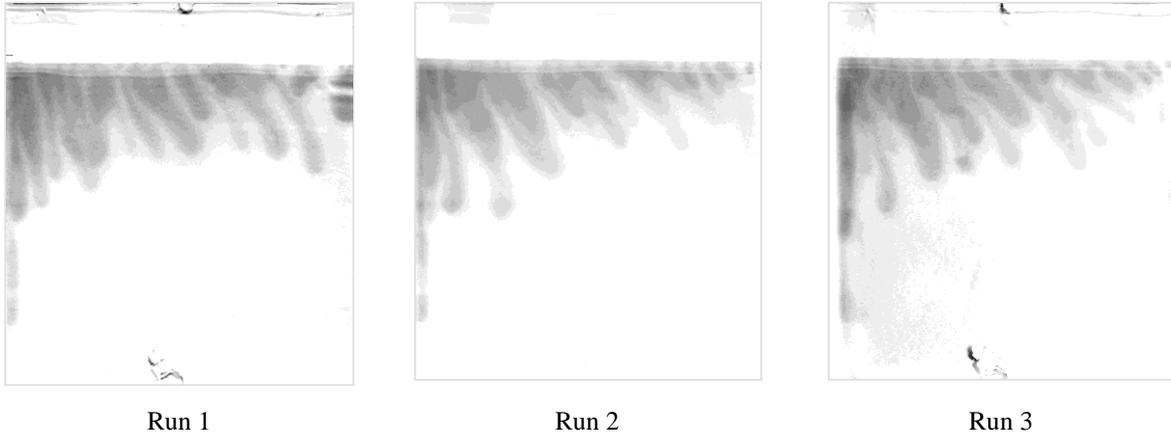

|  Run 1  |  Run 2  |  Run 3  |

**Figure 4.** Three runs for $CO_2$ introduced at 0.59 L/min to the left side of the cell at 85 min (CF, 1 mm aperture, vertically oriented flat glass)

The experimental investigations considered in our study can be divided into four cases, as presented in **Table 1.** In **Case A**, we consider the effect of the $CO_2$ injection point by varying the placement of the $CO_2$ injection point in the Hele-Shaw cell. Moreover, we also observe the effect of salinity by dissolving 1M $CaCl_2$ in the Control Fluid.

In order to observe the effect of initial cation concentration, we use $MgCl_2$ and $CaCl_2$ with varying concentrations in **Case B**, as shown in **Table 1.** The $Mg^{++}$ and $Ca^{++}$ cations are used in this study as they have received limited attention compared to numerous studies for Na+ cations. Furthermore, $Mg^{++}$ and $Ca^{++}$ cations form a significant fraction of the ions in worldwide saline aquifers [59]. Therefore, we also introduce a slight dipping ($\leq 1°$) to mimic a dipping aquifer by raising the right side of the Hele-Shaw cell slightly higher than the left.

Although **Cases A** and **B** consider homogeneity throughout the cell, $CO_2$ geological storage sites are heterogeneous. Barriers like calcite and shale layer are common forms of heterogeneity that can significantly dictate the density-driven natural convection mechanism [11]. Moreover, different configurations and geometries of the heterogeneities result in various convective flow directions and movements. Although several studies have introduced different heterogeneous patterns in the system [11,52–57] to observe and quantify the $CO_2$ dissolution-driven convection pattern, the effect of different salts in these heterogeneous systems is yet to be investigated. Therefore, in **Case C,** we add heterogeneities in the Hele-Shaw system by introducing regions with lower permeability values and observing the convection pattern change for solutions containing different salts. Moreover, despite the recognized importance of heterogeneous layers, the impact of fractures present in these heterogeneous formations has not been studied extensively [45,46]. Therefore, for **Case D**, we created narrow spacing inside heterogeneous formations to mimic the convective flow through fractured regions. The $CO_2$ flow rate in all the experiments in **Case B, C,** and **D** is kept constant at a flow rate of 0.59 L/min to remove any effect of injection rate.

**Table 1.** The experiments considered in this study and the investigated parameters

| Solution Used | $CO_2$ Injection Point | Case Number | Dipping Applied | Heterogeneity Introduced | Effect of Parameters Observed |
|---|---|---|---|---|---|
| *Control Fluid* (CF)[a] | Side | A1 | No | No | • Effect of the $CO_2$ injection point |
| CF | Middle | A2 | No | No | |
| 1 mole $CaCl_2$ dissolved in CF | Middle | A3 | No | No | • Effect of salinity |
| 1 mole $CaCl_2$ dissolved in CF | Side | B1 | Yes | No | • Effect of salt types with varying concentration |
| 2 mole $CaCl_2$ dissolved in CF | Side | B2 | Yes | No | |
| 1 mole $MgCl_2$ dissolved in CF | Side | B3 | Yes | No | |
| 2 mole $MgCl_2$ dissolved in CF | Side | B4 | Yes | No | |
| | | | | | • Effect of dipping |
| CF | Side | C1 | No | Yes | • Effect of Heterogeneity |
| 1 mole $CaCl_2$ dissolved in CF | Side | C2 | No | Yes | |
| 1 mole $MgCl_2$ dissolved in CF | Side | C3 | No | Yes | |
| 1 mole NaCl dissolved in CF | Side | C4 | No | Yes | |
| CF | Side | D1 | No | Yes | • Effect of Fracture |

[a]*Control Fluid* (CF) is prepared by mixing deionized 1 L of de-aired water with 0.0114% w/v BCG.

## 3. Results

### 3.1. Qualitative Visualization for Homogenous Cases

#### 3.1.1. Case A

We used the CF with varying injection points to test the effect of injection points in this section. Additionally, for **Case A3**, we dissolve 1 mole $CaCl_2$ in CF to investigate the impact of salinity. The detailed effect of different salts with varying concentrations will be discussed later.

**$CO_2$ introduced to the side of the cell (Case – A1)**

We introduce the gaseous $CO_2$ from the leftmost injection port for this case. The development of convection-driven fingering patterns is similar to what has been observed in previous studies [24,26,40,60]. $CO_2$ dissolution via diffusion was the only applicable mechanism at the initial stage of the visualization test, which we refer to as the $CO_2$ induction phase. During the induction phase, there is a uniform change in pH near the gas-water interface, indicating the diffusion of $CO_2$, where the gaseous $CO_2$ dissolves in water to form aqueous $CO_2$ and forms equilibrium with carbonic acid, undergoing the following reactions [59]:

$$CO_2(g) \rightarrow CO_2(aq) \qquad (3)$$

$$CO_2(aq) + H_2O \underset{}{\overset{K_1}{\rightleftarrows}} H^+ + HCO_3^- \qquad (4)$$

$$HCO_3^- \underset{}{\overset{K_2}{\rightleftarrows}} H^+ + CO_3^{--} \qquad (5)$$

$$H_2O \underset{}{\overset{K_3}{\rightleftarrows}} H^+ + OH^- \qquad (6)$$

During this period, the diffusive layer at the interface gradually expanded without any deformation or instability. We observed the formation of fingers at around 135 ± 15 seconds, demonstrating the end of the induction phase and onset of the convection process causing instability in the form of fingers, as shown in **Table 2**. It should be noted that we considered the time frame for which the fingers are clearly visible; therefore, the actual onset of convection should be slightly earlier since the earliest initiation of finger forming marks the onset of convective flow.

As time progressed, the fingers increased in width and depth, traveling downwards vertically. The fingers started merging as they grew in width, as evidenced by the decrease in the number of fingers and an increase in the wavelength (please refer to **Section 5.1** for the quantitative measurement). Moreover, as time progressed, we observed cell-scale convection, a phenomenon bringing fresh solution (solution not laden with $CO_2$) to the upper surface from the center, as shown by small new fingers forming at the gas-water interface. The formation of distinctive new fingers (also referred to as "nascent fingers") between established fingers can be seen as early as 27 minutes into the test. This phase is identified as the reignition phase, which lasts until the end of the experiment, also observed in other studies [26,40].

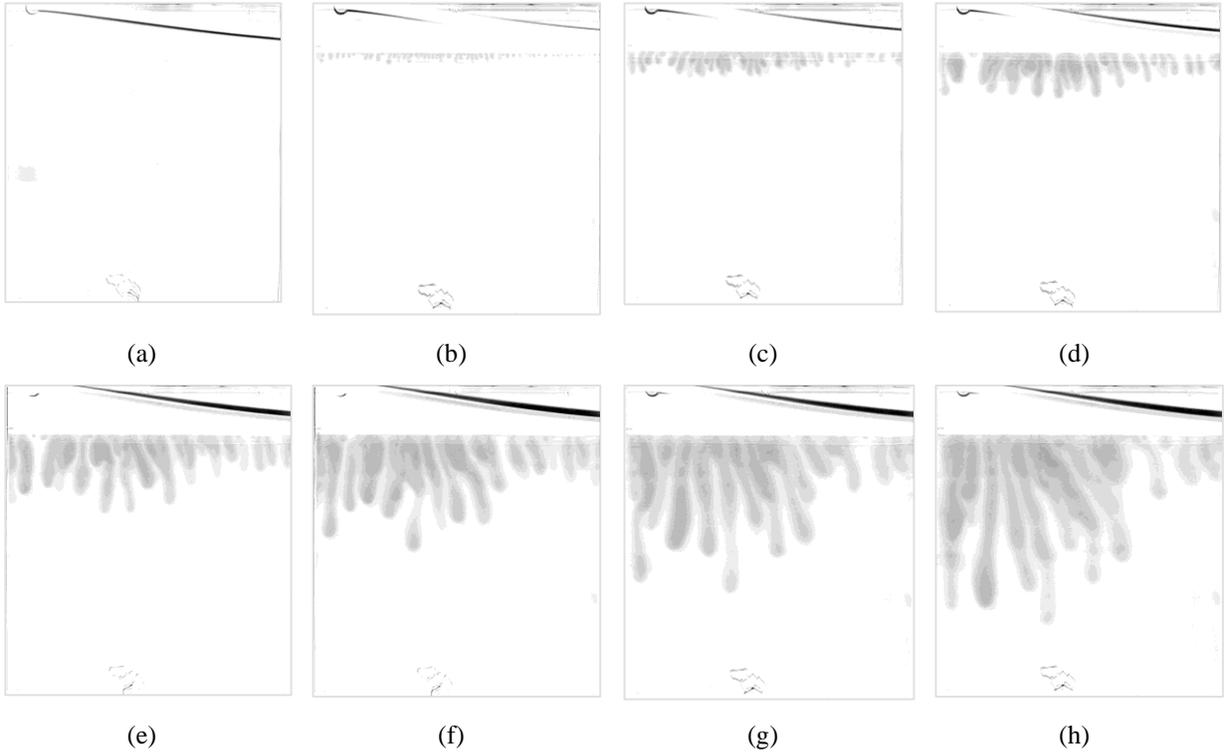

(a)      (b)      (c)      (d)

(e)      (f)      (g)      (h)

**Figure 5.** $CO_2$ introduced to the left side of the cell (CF, 1 mm aperture, vertically oriented flat glass) (a) Initial (no $CO_2$) (b) 9 minutes (c) 24 minutes (d) 36 minutes (e) 58 minutes (f) 85 minutes (g) 118 minutes (h) 154 minutes

The cell-scale convection also leads to the non-vertical movement (slight curve formed) made by the established fingers, which would otherwise move vertically due to density change and gravitational pull. The curved movement of the fingers can be observed as early as around 9 minutes into the visualization test (**Fig. 5, Frame 2 - 9 minutes**), which becomes more prominent as time progresses.

The whole area of the Hele-Shaw can be classified into two regions: active and inactive zones. The active zone can be referred to as the entire vertical region under the $CO_2$ injection point. In contrast, the region farther from the injection point can be identified as the inactive zone. As seen from **Fig. 5,** when $CO_2$ is injected into the left side of the cell, the area farther away from it, i.e., the inactive zone, has partially slowed finger formation. This difference can be attributed to the upward fluid flow due to cell-scale convection, which slows down the speed of the fingers. However, since the active zone has continuous $CO_2$ injection from the top, the effect of cell-scale convection in slowing down the finger flow is not as significant. Furthermore, due to the increasing difference in finger length between the active zone and inactive zone, we see an increase in the standard deviation of the finger length (refer to **Section 5.1** for a quantitative overview). Thus, injection near the boundary is clearly shown to have a different convective finger pattern from the injection at the middle of the cell, conducted in ours and several other studies [11,20,40].

### $CO_2$ introduced to the middle of the cell (Case - A2)

To observe the effect of different injection points, as a slight variation of the first base **Case A1** we changed the injection point for **Case A2** (**Fig. 6**). We observed the $CO_2$ induction phase followed by a convection period, like the first base case. Furthermore, we also observed similar cell-scale convection during this visualization test.

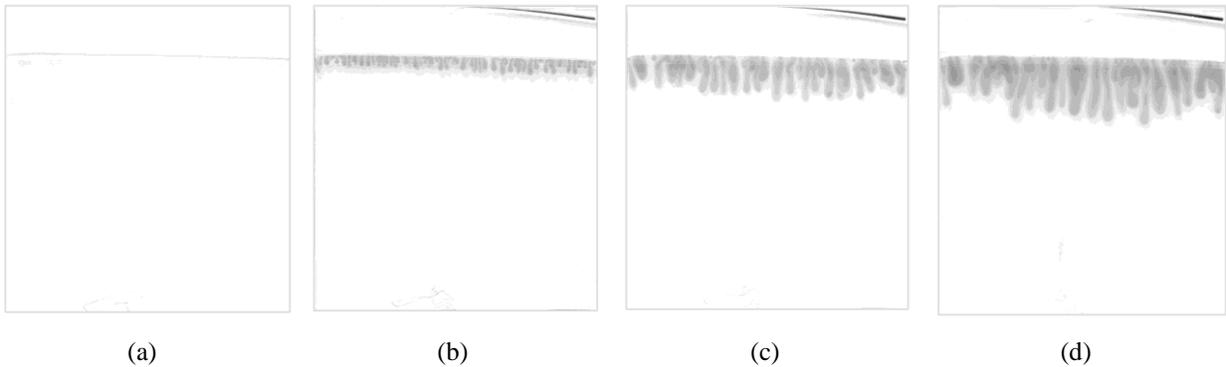

(a)    (b)    (c)    (d)

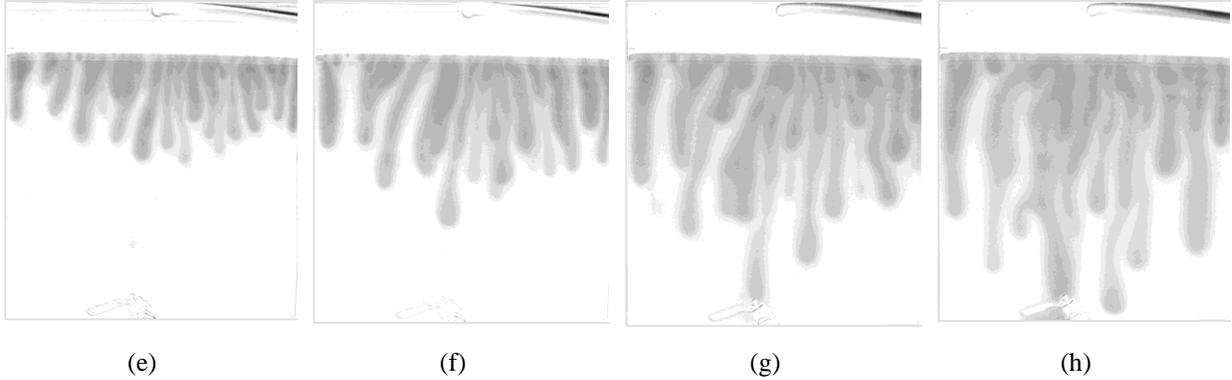

(e)          (f)          (g)          (h)

**Figure 6**. $CO_2$ introduced to the center of the cell (CF, 1 mm aperture, vertically oriented flat glass) (a) Initial (no $CO_2$) (b) 9 minutes (c) 24 minutes (d) 36 minutes (e) 58 minutes (f) 85 minutes (g) 118 minutes (h) 154 minutes

Changing the $CO_2$ injection point also affected the layout of the active and inactive zones. For the first base case with $CO_2$ injection at the extreme left side of the cell, the inactive zone was much farther away at the right side of the cell. However, by injecting $CO_2$ in the middle of the Hele-Shaw cell, the inactive zones (the left and right sides of the cell) were relatively close to the active zone. The dissolution was symmetrical around the hypothetical vertical central line. At similar experimental conditions, Taheri et al. [11] observed the convection fingers at the side of the boundary cell moving faster than the middle of the cell. They named this phenomenon the 'side boundary effect,' which is not observable in our study. This can be due to the smaller gap width (0.25 mm) or larger Hele-Shaw cell dimensions (500 cm × 500 cm) adopted in their study.

As seen in **Fig. 5**, the fingers in the inactive zone had slower finger formation compared to those in the active zone, leading to a high standard deviation with time. However, **Fig. 6** shows that despite the inactive zone having slowed finger formation, it is not as significant as the previous base case **A1**, also evidenced by the smaller standard deviation in finger length compared to the previous test (see **Section 5.1**). This visualization proves that injection points can impact finger evolution. In the field scale, injection nearby vertical no-flow boundaries may impact the convective finger formations, so they should be considered during $CO_2$ geological storage.

Onset time of convection, or the time at which the host phase experiences instability in the dissolution-driven convection flow, is an important aspect of safe $CO_2$ storage and risk assessment. A shorter onset time of convection is desired, as convection enhances the mixing of $CO_2$ in brine, allowing faster $CO_2$ movement away from the caprock downwards in the brine [40]. **Table 2** provides the onset time of convection and corresponding finger wavelength for the cases in **Case A. Table 2** shows that the injection point does not impact the onset time of convection, which is observed to be around 135 seconds (**Case A1** and **A2**). However, the convection initiates earlier in the presence of a salt ($CaCl_2$). This result agrees with Jiang et al. [61], who observed the onset time to be 130 s for pure water, 100 s for 0.25 wt.% saline water, and 20 s for 1.00 wt.% saline water.

Additionally, we observe the finger wavelength to be higher in **Case A3**, as the number of fingers formed is less in the presence of a salt. Furthermore, when comparing **Case A1** and **A2**, the wavelength at the onset of convection is higher when the injection point is at the side (**Case A1**). This difference can be attributed to the fact that during $CO_2$ injection at the side, the fingers at the inactive zone are not formed at the onset time of convection, thus reducing the number of fingers formed, resulting in a higher wavelength. The slower finger growth in the inactive zone for **Case A1** is also observable in **Fig. 5b**. We also considered the wavelength at 9 minutes, as shown in **Table 2. By that time,** all the fingers are visible and fully formed throughout the interface. At that period, the wavelength is almost similar for **Case A1**(4.55 mm) and **A2** (5.41 mm).

**Table 2.** Comparison and visual representation of the fingers at onset time of convection for Case A

| Experiment Case | Solution Used | Picture at Onset Time of Convection[a] | Time (sec)[b] | Wavelength (mm)[c] At Onset of Convection | At 9 minutes |
|---|---|---|---|---|---|
| Case A1 | CF | 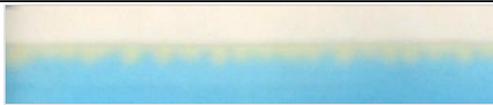 | 135 ± 15 | 5.3 | 4.55 |
| Case A2 | CF | 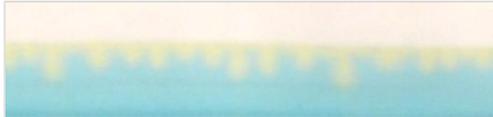 | 135 ± 15 | 3.56 | 5.41 |
| Case A3 | 1 mole CaCl$_2$ dissolved in CF | 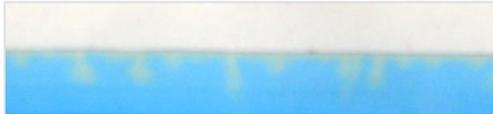 | 105 ± 15 | 6.33 | 6.83 |

[a] The region under the injection point is cropped and edited for better visibility
[b] For the onset time of convection, we consider the time for which the fingers are clearly visible
[c] Wavelength is the ratio of the length of interface and the number of fingers for that particular interval

**1 mole CaCl$_2$ dissolved in Control Fluid – CO$_2$ introduced to the middle of the cell (Case A3)**

The presence of salt stabilizes the convective dissolution of CO$_2$, as predicted in the theoretical work by Loodts et al. [39] and also shown experimentally in Thomas et al. [40]. By adding salt or increasing salt concentration, the solubility of CO$_2$ decreases, consequently reducing the density gradient between the fresh solution and the CO$_2$ mixed solution. Moreover, the system is also stabilized by the increasing viscosity of the solution caused by higher concentrations of salt [39,62]. Since the formation and development of convection-driven flow depends on density-driven gravitational instabilities, the convection process is slowed down in a stable system [20,39,40]. This is also evident from our visualization tests, as seen in **Fig. 7,** and our quantitative measurements (**Section 5**).

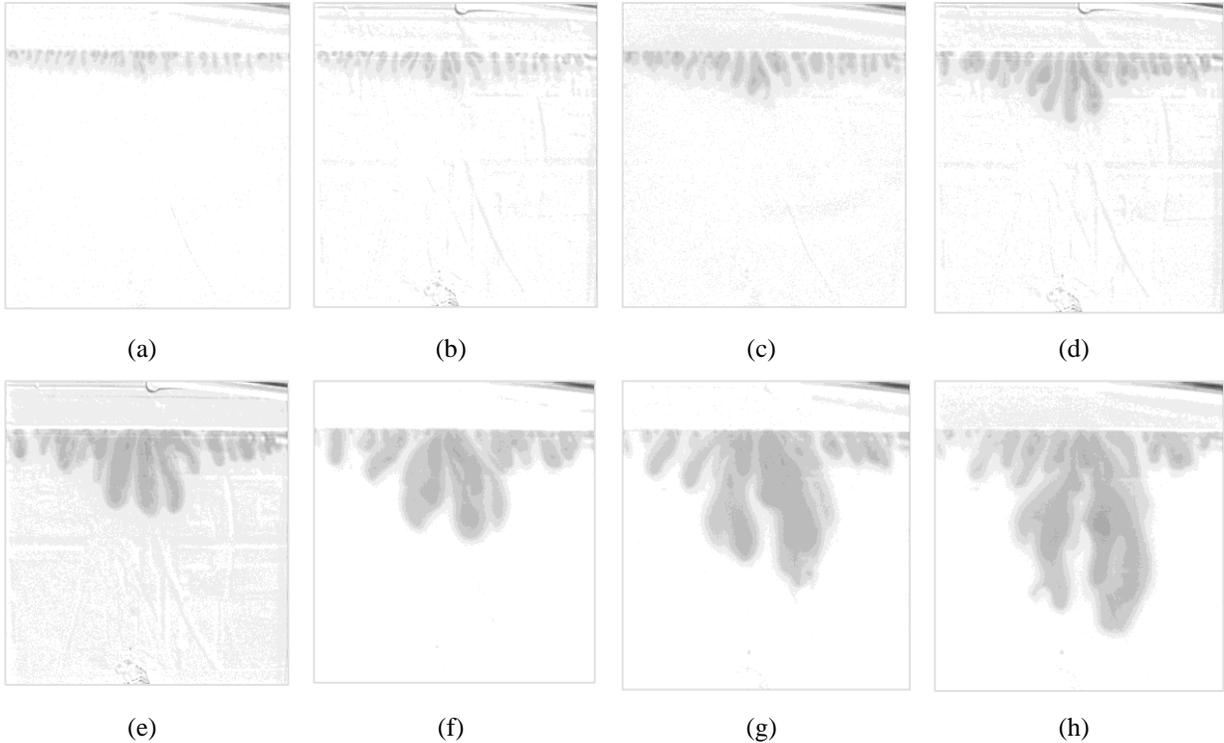

**Figure 7**. $CO_2$ introduced to the center of the cell (1 mole $CaCl_2$ dissolved in CF- 1 mm aperture, vertically oriented flat glass) (a) 9 minutes (b) 15 minutes (c) 24 minutes (d) 36 minutes (e) 58 minutes (f) 85 minutes (g) 118 minutes (h) 154 minutes

Although the symmetry around the hypothetical vertical line is maintained, the presence of $CaCl_2$ severely hindered the finger growth rate in the inactive zones, resulting in a large standard deviation throughout the test, as seen in **Fig. 7**. Moreover, we also observed a reduction in the number of fingers formed, the maximum being 38 at 540 seconds. Unlike the previous base cases where fingers were formed close to one another, the fingers formed in this case were placed at a minimal distance from each other (can be observed in **Fig. 7, Frame 2 -15 minutes**). The visualization tests show the importance of knowing the constituents and their composition level for a saline aquifer beforehand to predict the expected $CO_2$ storage behavior.

### 3.1.2. Effect of salt concentration on dissolution (Case B)

The dissolved chemical reactants in saline aquifers can affect the convective dynamics of $CO_2$ dissolution, as various geochemical reactions may occur between the dissolved acidic $CO_2$ and salts in the brine. Different reactants can trigger a change in the density stratification, modifying the nonlinear finger instability development [38]. In addition, the reservoir's physicochemical characteristics can determine whether the chemical reactions can enhance or retard the $CO_2$ dissolution process. In general, for A+B→C chemical reactions, the reaction product's influence on local density change with respect to the initially dissolved reactants determines the intensity of convective dissolution [38,39,60]. Although $CO_2$ solubility in NaCl solutions has been extensively studied, other ionic solutions like $MgCl_2$ and $CaCl_2$ have not received such attention [4,39,40,59,63–65]. The pore fluid in saline aquifers contains different compositions of ionic solutions, with common components including Calcium, Magnesium, Potassium, Iron,

Chloride, Sulphate, etc., which leads us to investigate the $CO_2$ dissolution-driven convection phenomenon in $CaCl_2$ and $MgCl_2$. [59,66]. The presence of different minerals in sandstone formations has a direct effect on $CO_2$ mineral trapping and also plays a role in altering reservoir transport properties [59,67]. For example, the presence of Anorthite ($CaAl_2Si_2O_8$), a common Ca-bearing feldspar, in sedimentary rock aids in faster reaction kinetics and higher $CO_2$ reactivity, thus enhancing the reservoir rock porosity as well [67]. We refer our readers to the review of Silva et al. [59] for the detailed geochemical aspects of $CO_2$ sequestration in deep saline aquifers.

As observed by Liu et al. [66], in the presence of salt ions like NaCl, $CaCl_2$, or $MgCl_2$ in pore fluid, water molecules bind with the "solvates," leaving less water for $CO_2$ dissolution. The precipitation reactions for $Ca^{2+}$ and $Mg^{2+}$ ions result in the formation of Calcite ($CaCO_3$) and Magnesite ($MgCO_3$), respectively, as shown in **Equations 7 and 8** [59].

$$Ca^{2+}(aq) + CO_3^{2-}(aq) \rightarrow CaCO_3(s) \tag{7}$$

$$Mg^{2+}(aq) + CO_3^{2-}(aq) \rightarrow MgCO_3(s) \tag{8}$$

Even though most theoretical benchmarks and simulation studies consider ideally horizontal geological storage sites, it is exceedingly rare in real-life scenarios. Most of the storage sites have a formation dip angle, thus making it another key factor of consideration for safely storing $CO_2$ on subsurface geological sites [42–44]. To visualize the effect of different salts with varying concentrations with a minimal dipping angle, we introduced $CO_2$ to the left side of the cell at a flow rate of 0.59 L/min, with the right side of the Hele-Shaw cell placed about 1° higher than the left side for the experiments in **Case B**. This allows us to inspect the $CO_2$ convective dissolution in regions farther from the injection point and the effect of dipping. Furthermore, the qualitative findings from the experiments allow investigation of the changes due to the presence of different salts on saline aquifers. We prepare the solution by dissolving salt with different concentrations in the CF for the visualization tests.

**$CaCl_2$ dissolved in Base Case Fluid – $CO_2$ introduced to the side of the cell (Case B1 and Case B2)**

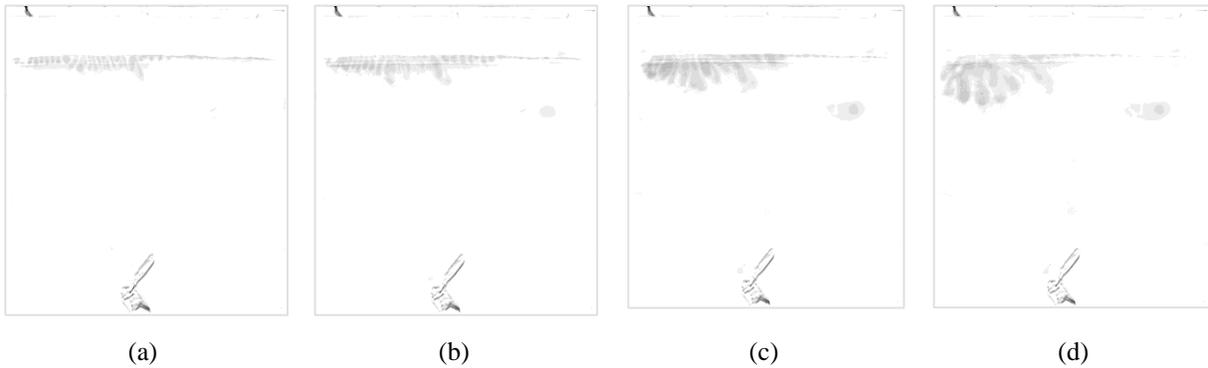

(a)      (b)      (c)      (d)

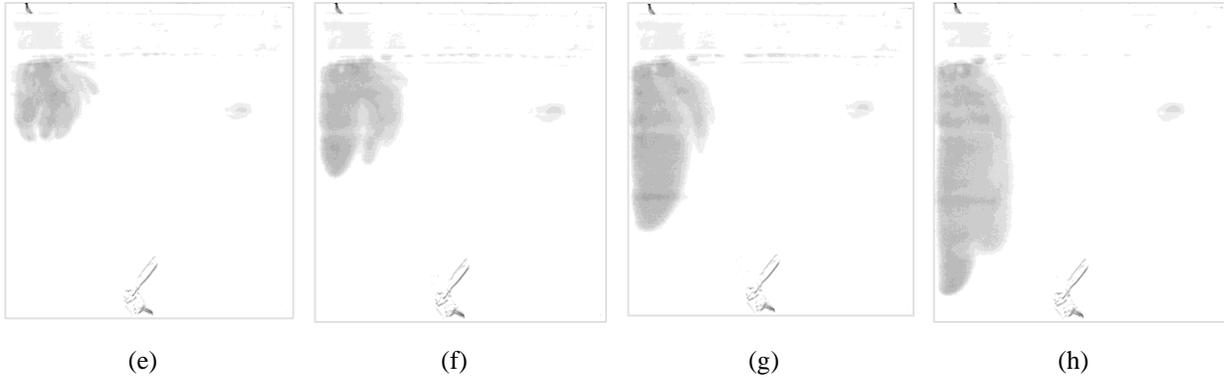

(e)　　　　　　　　(f)　　　　　　　　(g)　　　　　　　　(h)

**Figure 8**. $CO_2$ introduced to the side of the cell at 0.59 L/min (1 mole $CaCl_2$ dissolved in CF- 1 mm aperture, vertically oriented flat glass, **Case B1**) (a) 9 minutes (b) 15 minutes (c) 24 minutes (d) 36 minutes (e) 58 minutes (f) 85 minutes (g) 118 minutes (h) 154 minutes

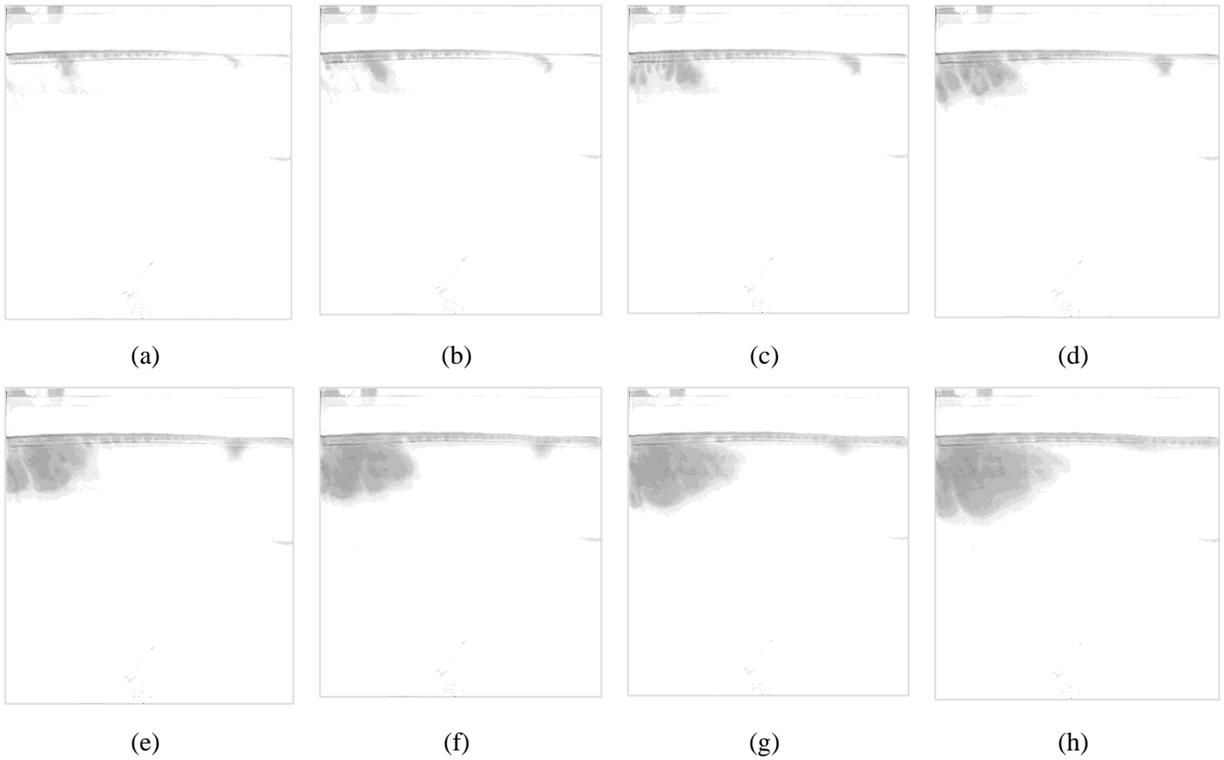

(a)　　　　　　　　(b)　　　　　　　　(c)　　　　　　　　(d)

(e)　　　　　　　　(f)　　　　　　　　(g)　　　　　　　　(h)

**Figure 9**. $CO_2$ introduced to the side of the cell at 0.59 L/min (2 mole $CaCl_2$ dissolved in CF- 1 mm aperture, vertically oriented flat glass, **Case B2**) (a) 9 minutes (b) 15 minutes (c) 24 minutes (d) 36 minutes (e) 58 minutes (f) 85 minutes (g) 118 minutes (h) 154 minutes

The effect of the $CO_2$ injection point near the boundary and the dipping angle is observable by comparing **Case A3**, **Case B1, and B2.** For **Case B1 and B2**, the fingers formed in the vicinity of the boundary can't move freely due to restricted space and merge with the nearby fingers, resulting in bigger oval-shaped finger formations, as seen in **Fig. 8 and 9**. Moreover, due to the dipping angle, with the right side of the cell being placed higher, the finger formation on the right side is severely impacted. The finger formed on the right side moves laterally and mixes with the left side fingers, aiding in the oval-shaped finger formation. The formation of nascent-fingers is observed, and the cycle of lateral finger movement and finger-merging continues throughout the experiment. Whereas, without any boundary effect, as represented by $CO_2$ injection in the middle (**Case A3**), the coalesced fingers in the active zones protrude in different directions rather than forming a large oval-shaped finger formation.

The effect of increased salt concentration is visualized in **Case B2** by changing the concentration of $CaCl_2$ to 2M while keeping all the other parameters the same as in **Case B1.** The solution in **Case B2** has a higher $CaCl_2$ concentration, resulting in a lower $CO_2$ solubility. Moreover, the formation and development of convective-driven flow are slowed in this case, as shown in **Fig. 9**. The vertically downward finger movement is slowed significantly, alluding to the reduced impact of gravity-driven convectional instabilities.

**Effect of MgCl$_2$ with different concentration (Case B3 and Case B4)**

**Fig. 10 and 11** show that the solution with $MgCl_2$ has higher CO2 convective flow than CaCl2, which is evident from a higher pH depressed area (indicating dissolved $CO_2$) in the pictures. The $CO_2$ dissolved area indicated by the pH-depressed region follows the order of $CaCl_2$ 2M solution < $MgCl_2$ 2M solution < $CaCl_2$ 1M solution < $MgCl_2$ 2M solution. Detailed quantitative analysis of the pH depressed zone for the salts is provided in **Section 5**. Besides that, we observe the same effect of dipping angle on $CO_2$ dissolution in the presence of $MgCl_2$. Lateral mixing is observed in both lower and higher concentrations of $MgCl_2$. However, the vertical movement of the finger is considerably slowed down at 2M $MgCl_2$ solution, as shown in **Fig. 11**.

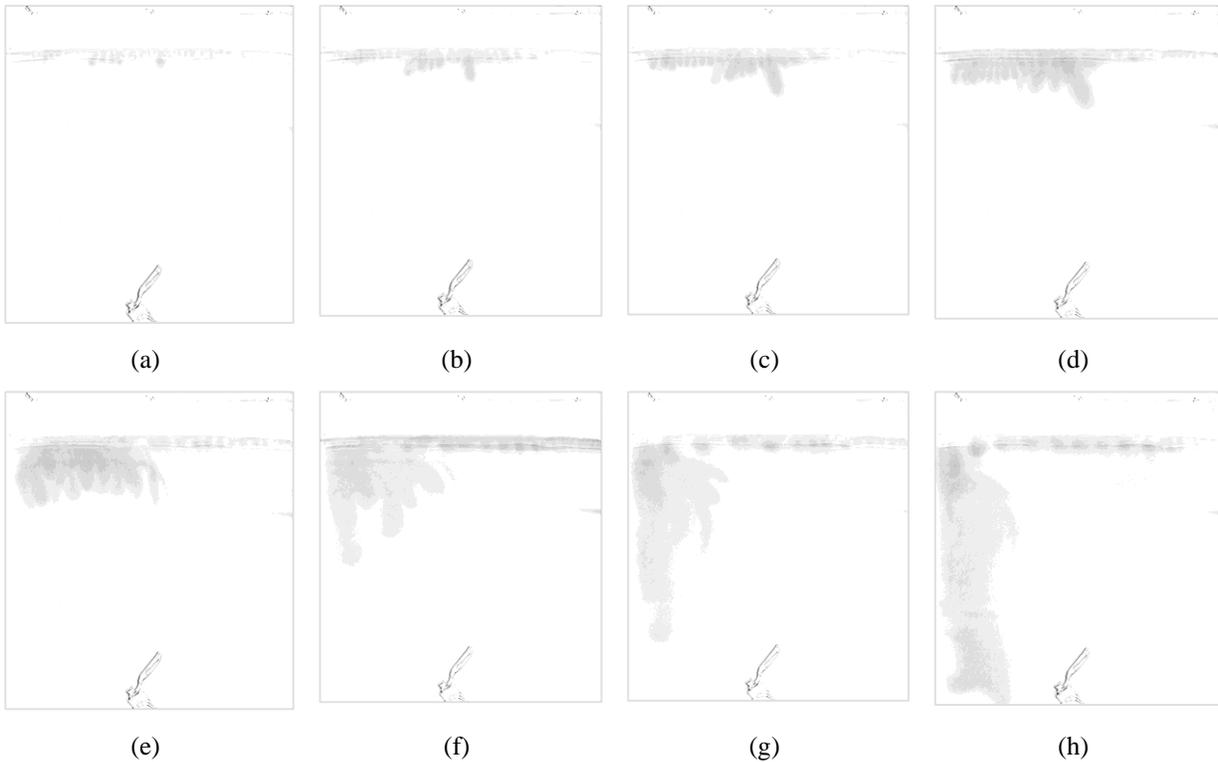

(a)　　　　　　　(b)　　　　　　　(c)　　　　　　　(d)

(e)　　　　　　　(f)　　　　　　　(g)　　　　　　　(h)

**Figure 10**. $CO_2$ introduced to the side of the cell at 0.59 L/min (1 mole $MgCl_2$ dissolved in CF- 1 mm aperture, vertically oriented flat glass, **Case B3**) (a) 9 minutes (b) 15 minutes (c) 24 minutes (d) 36 minutes (e) 58 minutes (f) 85 minutes (g) 118 minutes (h) 154 minutes

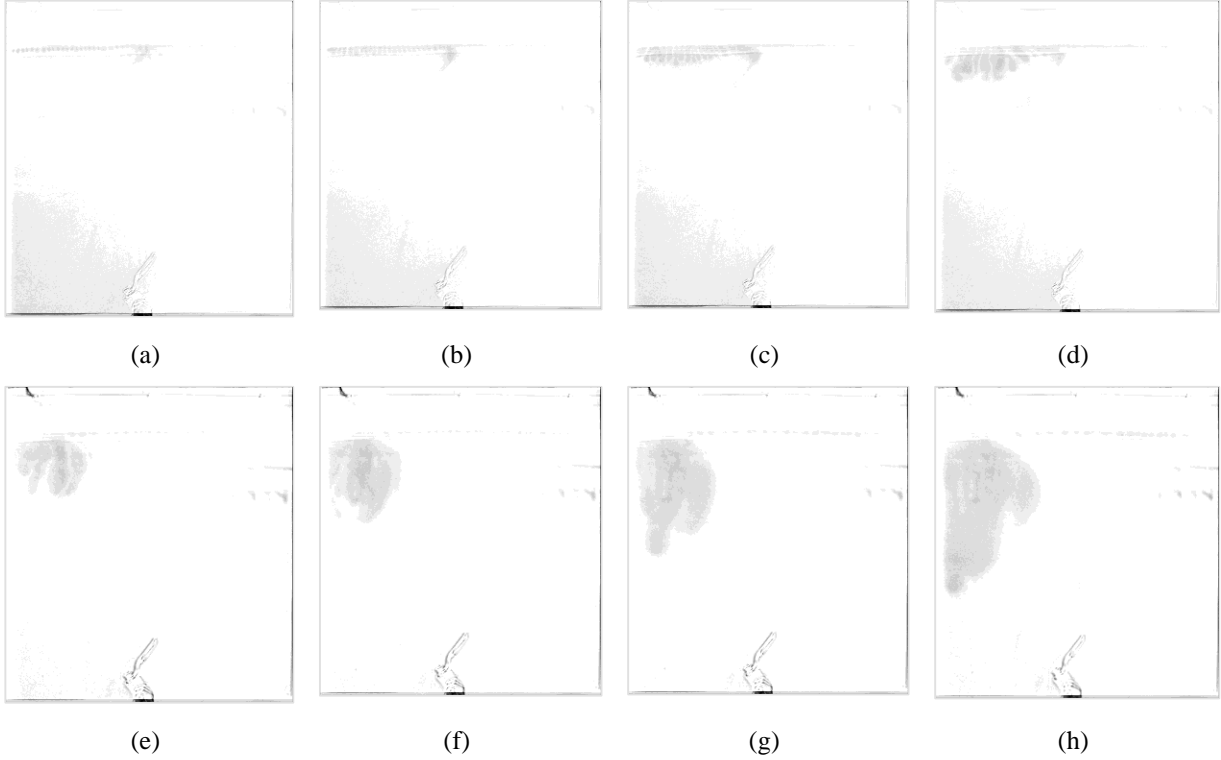

(a) (b) (c) (d)

(e) (f) (g) (h)

**Figure 11**. $CO_2$ introduced to the side of the cell at 0.59 L/min (2 mole $MgCl_2$ dissolved in CF- 1 mm aperture, vertically oriented flat glass, **Case B4**) (a) 9 minutes (b) 15 minutes (c) 24 minutes (d) 36 minutes (e) 58 minutes (f) 85 minutes (g) 118 minutes (h) 154 minutes

### 3.2. Qualitative Visualization for Effect of Permeability Heterogeneity (Case C and D)

All the storage sites have some form of vertical and horizontal permeability heterogeneities introduced by the presence of calcite layers or shales, which play a major role in dictating fluid flow [11,68]. Different convective-flow patterns (gravity fingering, channeling, or dispersion) can be observed based on the heterogeneity medium. Depending on the heterogeneity parameters, the average $CO_2$ mass flux at the top boundary can reach different constant values after decreasing at the initial stage and then increasing, as shown by Ranganathan et al. [69].

In this section, we investigate the effect of heterogeneities on convective behavior by setting up multiple 0.05 mm thick graphite strips inside the Hele-Shaw cell, thus reducing the aperture to 0.05 mm for these regions. The length and height of these graphite strips are 50 mm and 12.7 mm, respectively. For better visualization of the graphite strips, we have highlighted them in the figures. The permeability for the region with graphite strip was $2.083 \times 10^{-8}\,m^2$, which was calculated by assuming that the subsection behaves like an independent miniature Hele-Shaw cell with an aperture size of 0.05 mm by using **Eq. 1**. Additionally, to determine the effective permeability of the Hele-Shaw cell, we consider the permeability of both the regions with and without the graphite strips and calculate the weighted average based on the area. The updated porosity is calculated as the ratio of available volume inside the Hele-Shaw to the total volume and is found to be 0.948. The effective permeability for the medium with heterogeneities is calculated as 7.68 × $10^{-8}$ m$^2$, with a corresponding Rayleigh number of 40663, which is lower than the permeability and Ra number obtained for our homogeneous cases by 7.72% and 2.68%, respectively. Although the permeability calculated using this method may not strictly represent the actual permeability due to the associated boundary effects, the overall trend remains valid.

### 3.2.1. Effect of Heterogeneity (Case C1, C2, C3, and C4)

**Fig. 12** displays the $CO_2$ dissolution behavior in the presence of heterogeneous barriers. The downward velocity of the fingers is slowed down along with decreased $CO_2$ dissolution, as evidenced by the smaller pH-depressed region compared to the homogenous case (**Fig. 3**). Furthermore, as shown in **Fig. 12 (e-h),** upon passing the first set of heterogeneous layers, rather than traveling downwards vertically, the $CO_2$ finger move in a curved manner towards the side of the heterogenous strips in the second layer. The curved finger travel alludes to a preferential $CO_2$ movement path, which can vary on the geometry of the heterogeneity. The heterogeneity also promotes flow channeling, resulting in a large region that CO2 does not encounter.

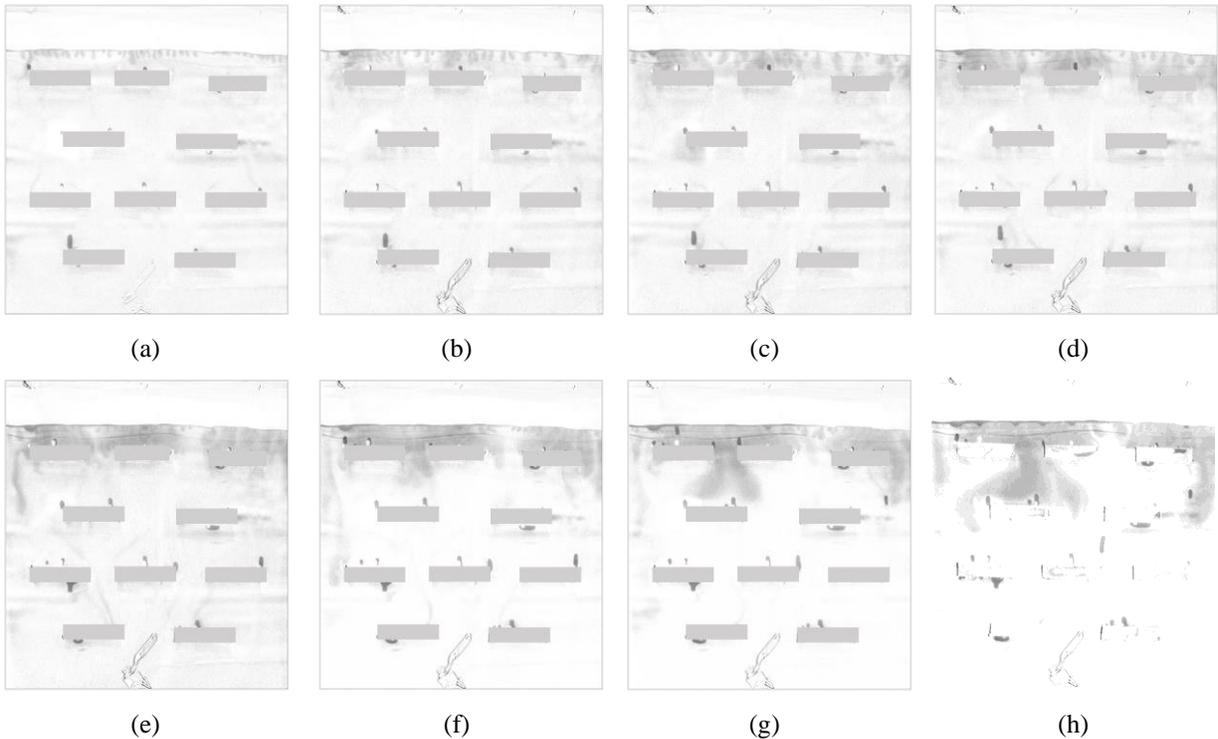

(a) (b) (c) (d)

(e) (f) (g) (h)

**Figure 12**. CO$_2$ introduced to the side of the cell at 0.59 L/min (CF, 1 mm aperture, vertically oriented flat glass with heterogeneous layers, **Case C1**) (a) 9 minutes (b) 15 minutes (c) 24 minutes (d) 36 minutes (e) 58 minutes (f) 85 minutes (g) 118 minutes (h) 154 minutes

**Fig. 13, 14,** and **15** represent the $CO_2$ dissolution behavior in the presence of 1M $CaCl_2$ solution, 1M $MgCl_2$ solution, and 1M NaCl solution, respectively. For NaCl and $CaCl_2$, as the convection is slower than CF, the lateral merging of $CO_2$ fingers is more dominant once it encounters the heterogeneous barrier. However, as shown in **Fig. 14**, for $MgCl_2$, the vertical movement is more dominant compared to both NaCl and $CaCl_2$. This prevalent vertical finger travel for 1M $MgCl_2$ solutions can be attributed to faster $CO_2$ dissolution in the $MgCl_2$ solution than in NaCl and $CaCl_2$ solutions. The more rapid dissolution allows the fingers to quickly pass the heterogeneous barriers, leaving less time for lateral mixing.

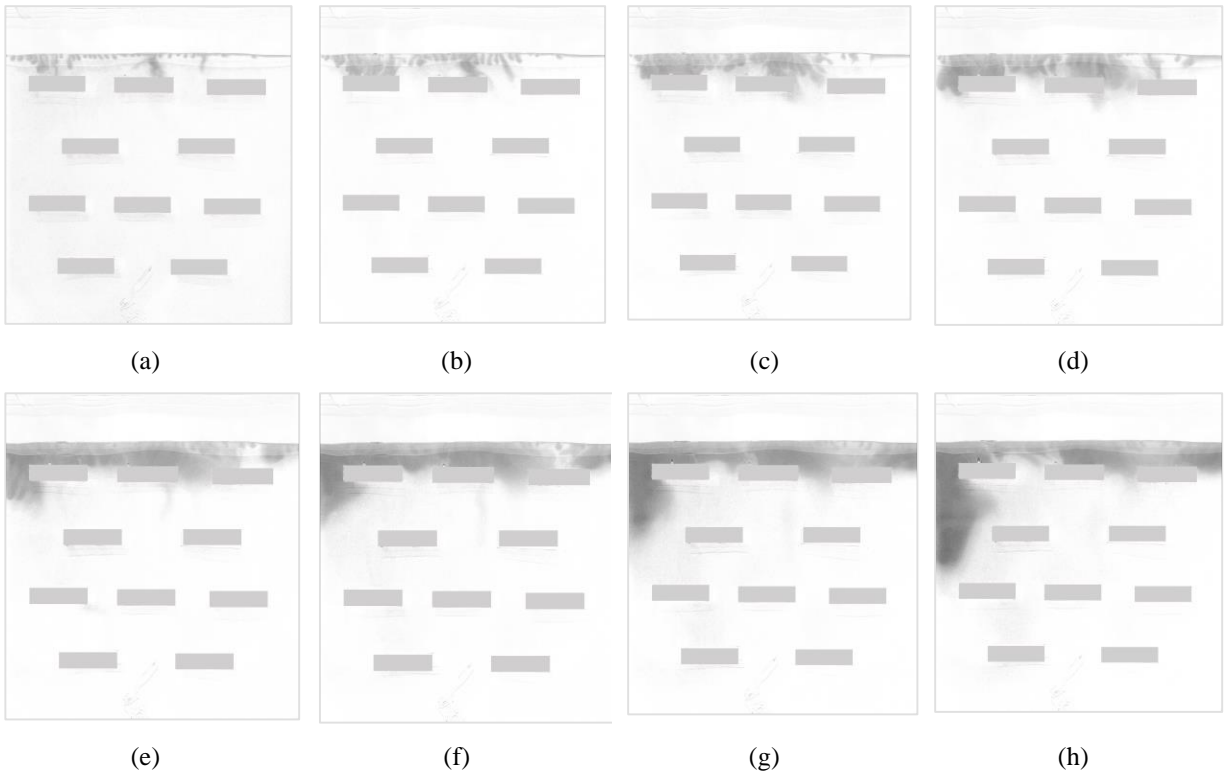

(a) (b) (c) (d)

(e) (f) (g) (h)

**Figure 13**. $CO_2$ introduced to the side of the cell at 0.59 L/min (1 mole $CaCl_2$ dissolved CF- 1 mm aperture, vertically oriented flat glass with heterogeneous layers, **Case C2**) (a) 9 minutes (b) 15 minutes (c) 24 minutes (d) 36 minutes (e) 58 minutes (f) 85 minutes (g) 118 minutes (h) 154 minutes

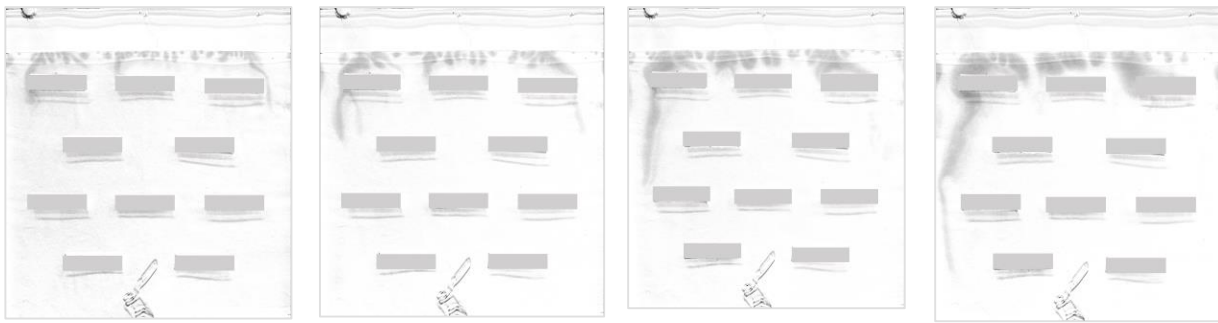

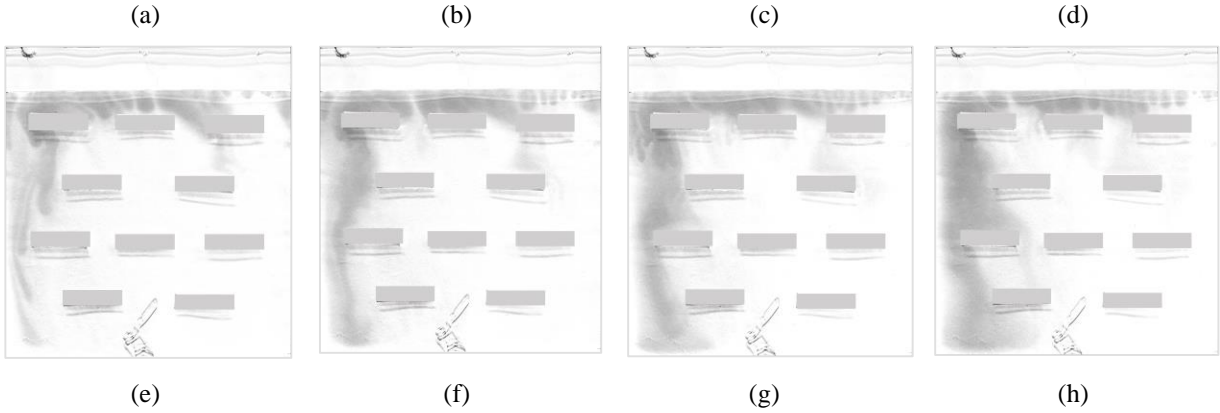

**Figure 14**. CO₂ introduced to the side of the cell at 0.59 L/min (1 mole MgCl₂ dissolved CF- 1 mm aperture, vertically oriented flat glass with heterogeneous layers, **Case C3**) (a) 9 minutes (b) 15 minutes (c) 24 minutes (d) 36 minutes (e) 58 minutes (f) 85 minutes (g) 118 minutes (h) 154 minutes

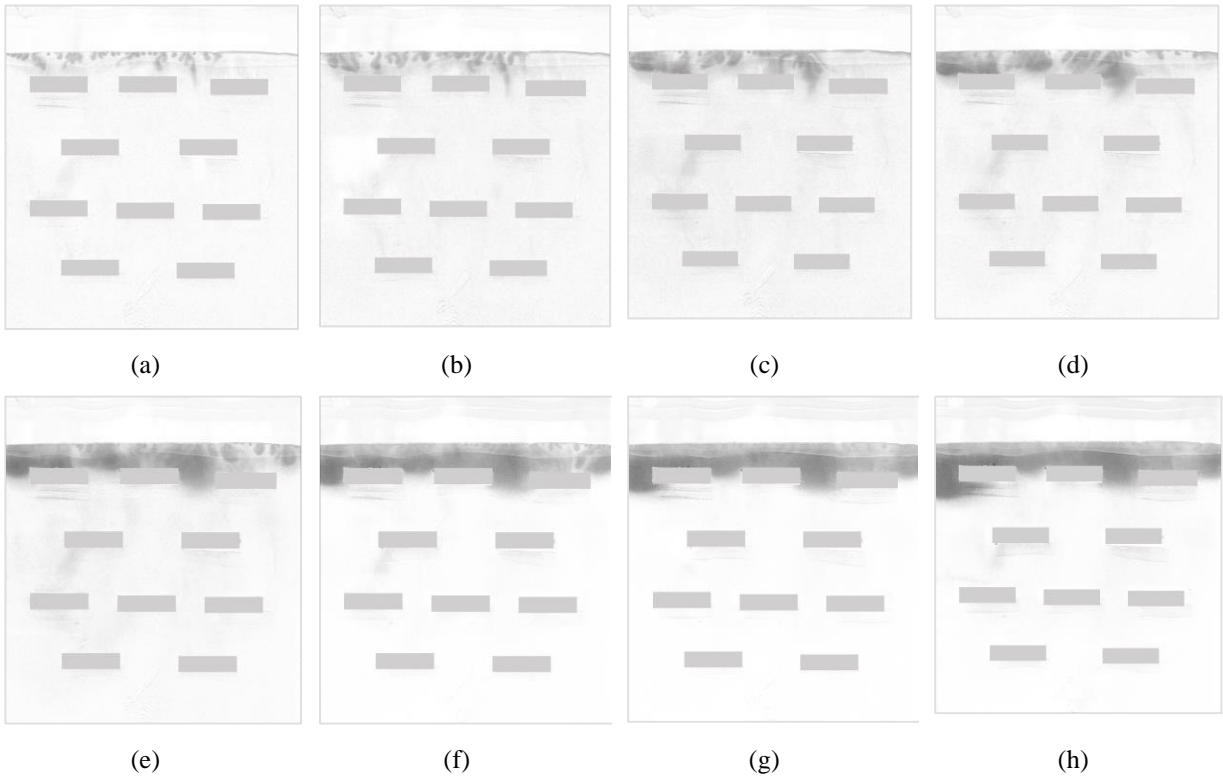

**Figure 15**. CO₂ introduced to the side of the cell at 0.59 L/min (1 mole NaCl dissolved CF- 1 mm aperture, vertically oriented flat glass with heterogeneous layers, **Case C4**) (a) 9 minutes (b) 15 minutes (c) 24 minutes (d) 36 minutes (e) 58 minutes (f) 85 minutes (g) 118 minutes (h) 154 minutes

### 3.2.2. Effect of Fractures in Heterogeneous Layers (Case D1)

Information about natural fractures and their interaction with $CO_2$-acidified brine in the subsurface can play a crucial role in assessing the $CO_2$ storage parameters. Compared to the surrounding rock matrix, natural fractures in the geothermal reservoirs have a higher permeability region [70]. Fractures also have high conductivity, and the presence of an interconnected fracture system can provide potential $CO_2$ escape routes during storage by promoting the increased spatial spreading of $CO_2$ plume, thus allowing $CO_2$ migration pathway through the cap rocks to neighboring aquifers or surfaces [50,51]. Moreover, due to the low capillary pressure in the fractures (typical orders of magnitude lower than the neighboring rock matrix), capillary forces can prevent $CO_2$ from invading the rock matrix. This is significantly detrimental to the storage potential of the aquifer since most of the pore volume is accounted for by the matrix space. Fractures in storage sites can show a great variety in their morphology, permeability, roughness, and connectivity [71]. In real-life scenarios with more complex pore-geometry and increased heterogeneity, dissolution patterns will be non-uniform with faster channeling [46,72–74]. Moreover, with an increase in the fracture length and density, the connectivity of a fracture network will increase, resulting in a higher network permeability, as concluded in the work of Rossen et al. [75].

Despite the recognized importance of heterogeneous layers, the impact of fractures present in these heterogeneous formations on the convective-flow pattern has not been visually studied extensively [45,46]. Therefore, we study the effect of fractures in heterogeneous regions by creating thin slits (4 mm spacing at 45° angle) between the graphite shims, as shown in **Fig. 16**. **Fig. 16** shows that the $CO_2$ spatial spreading is much faster due to the fractures compared to cases without fractures in heterogeneity (**Case C1-4**). This can be attributed to an increase in effective permeability in the system. The regions underneath the top fracture layer show the point of the $CO_2$ dissolution path, as new $CO_2$ fingers are visible underneath them in **Fig. 16, Frame 2 -15 minutes.** Alluding to the channeling effect, a preferential dissolution path is clearly visible around an hour into the experiment (**Fig. 16, Frame 5**).

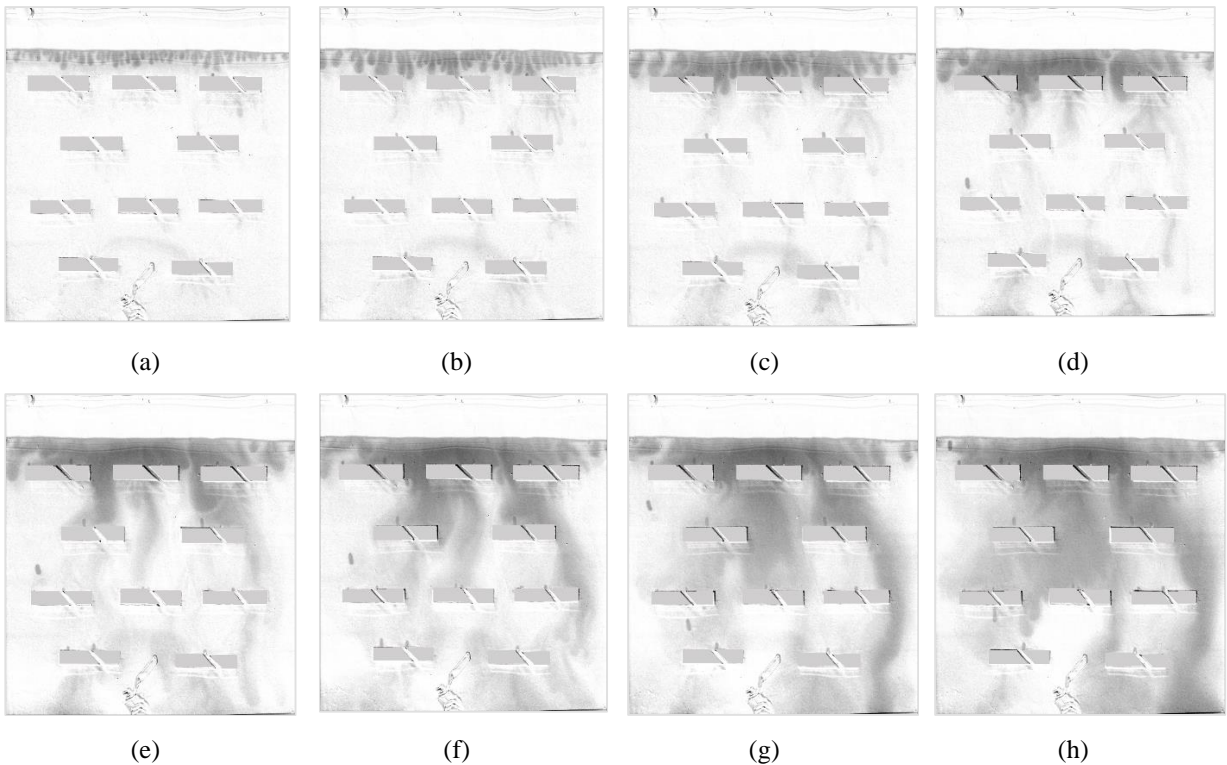

(a) (b) (c) (d)

(e) (f) (g) (h)

**Figure 16**. $CO_2$ introduced to the side of the cell at 0.59 L/min (CF- 1 mm aperture, vertically oriented flat glass with fractured heterogeneous layers, **Case D1**) (a) 9 minutes (b) 15 minutes (c) 24 minutes (d) 36 minutes (e) 58 minutes (f) 85 minutes (g) 118 minutes (h) 154 minutes

## 4. Quantitative measurements

The number of fingers formed, average finger length (mm), average wavelength (mm), and standard deviation of finger length are the parameters that characterize the finger evolution dynamics. Finger length is calculated as the vertical distance from the interface to the tip of the finger. In contrast, the average wavelength is the interface length divided by the number of fingers. It should be noted that fingers with different tips at the end moving in different directions are considered separate fingers in this study.

Before presenting the results of our experimental investigations, we perform a repeatability test for the quantitative data (refer to **Fig. 4**). As shown in **Fig. 17**, for the three runs, we calculate the average of the finger parameters and consider the deviation with a 95% confidence interval. Although we observed a slight difference in the number of fingers formed initially, the deviation becomes negligible in the middle and later parts of the experiment. We also observe a deviation from the mean finger length value in the latter part of the experiment. This deviation in the early part of the experiment can be due to the uncertain nature of convective finger merging and flow patterns as the finger grows. The acceptable level of deviation observed in **Fig. 17** shows the repeatability of our experiments. However, the convective flow pattern is sensitive to the experimental control parameters (dipping angle, injection orientation, the solution used, etc.) and must be followed with caution to ensure similar quantitative results.

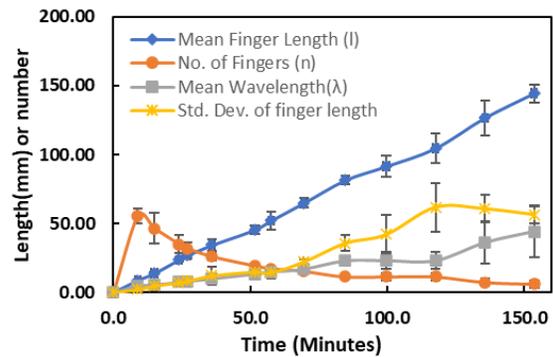

**Figure 17.** Quantitative data of the finger length parameters for the repeatability experiments.

### 4.1. Effect of injection point, $CO_2$ flow rate, and presence of salt

The image analysis for the base cases (**Fig. 18 a and b**) reveals average finger length having a similar trend, which denotes that the average finger length formation is independent of the injection point. However, **Fig 18 (b)** shows higher values of average finger length which can be attributed to the high $CO_2$ injection speed leading to faster finger formations. Moreover, the number of fingers formed remains similar, regardless of the $CO_2$ injection speed and injection point. However, the standard deviation of finger length is higher when $CO_2$ is injected into the left side of the cell. This is because the fingers underneath the $CO_2$ injection point quickly grew in size while the area on the right side observed less $CO_2$ dissolution, causing a higher imbalance in finger size.

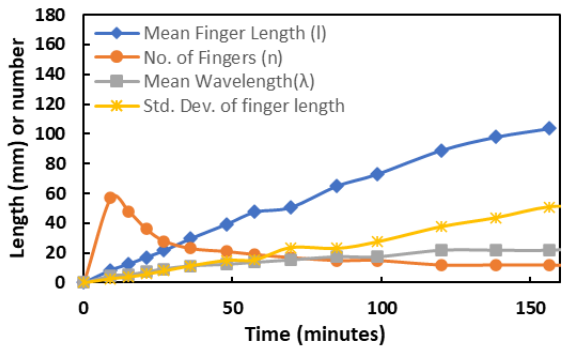
(a)

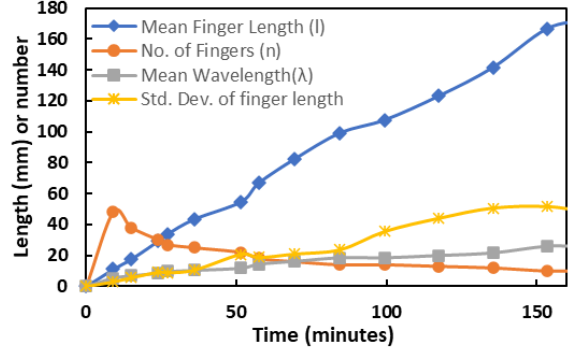
(b)

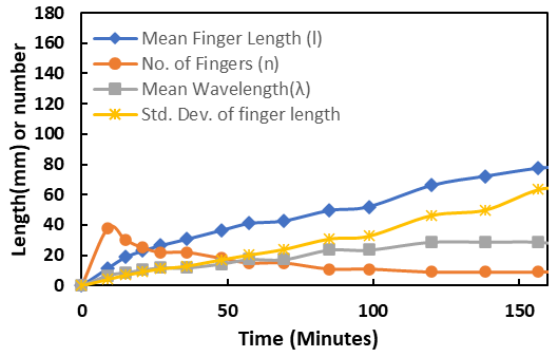
(c)

**Figure 18.** Number of fingers formed, wavelength, and finger length parameters for (a) Case A1 - $CO_2$ introduced to the side of the cell and (b) Case A2 - $CO_2$ introduced to the center of the cell (c) Case A3- 1 mole $CaCl_2$ dissolved in CF - $CO_2$ introduced to the center of the cell

The effect of salinity on finger evolution can be observed in **Fig. 18 (c)**. The effect of convection is attenuated due to the presence of $CaCl_2$, resulting in less finger interaction than in the control fluid. This is evident by the reduced number of fingers formed. Similar effects of salt were observed in other studies [39–41]. A higher standard deviation of finger length is also observed, which is due to the significant difference in finger size between the active zone (area under the $CO_2$ injection point) and the inactive zones (area at the side of the Hele-Shaw cells) (refer to **Section 3.2** for detailed explanation).

### 4.2. Effect of different salts with varying concentration

To illustrate the effect of the presence of different salts with varying concentrations, the pH-depressed region for **Case B1-B4** is calculated and compared with the $CO_2$ dissolution in the CF. **Fig. 19** shows the results normalized with respect to the pH-depressed area for the CF. $CO_2$ is injected at the side of the Hele-Shaw cell with a flow rate of 0.59 L/min for all the experiments, avoiding the impact of the injection point and flow rate on the convective-dissolution pattern. It should be noted that a higher pH-depressed area refers to more $CO_2$ dissolution. As seen in **Fig. 19**, higher concentration leads to slower $CO_2$ dissolution for the same salt type, which is in perfect agreement with other studies [20,39,40]. The $CO_2$ dissolved area indicated by the pH-depressed region follows the order of $CaCl_2$ 2M solution < $MgCl_2$ 2M solution < $CaCl_2$ 1M solution < $MgCl_2$ 2M solution. At the beginning part of the experiment, all the salt types have almost similar dissolution areas. It can be due to the impact of minimal convection-driven flow at the beginning of the experiment. Therefore, it can be safe to assume that the effect of salinity is not significant throughout the induction phase and for a small period after the onset of convection.

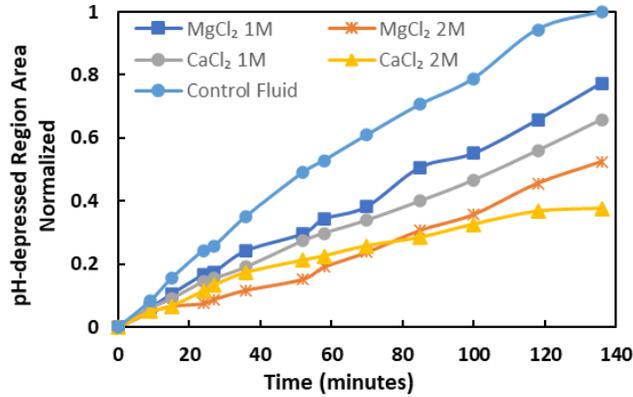

**Figure 19.** Area of the pH-depressed region (normalized) for different salts with varying concentration

The downward progression of fingers (also referred to as 'mixing length' by Thomas et al. [40]) is another factor worthy of investigation as it relates to the depth of investigation during $CO_2$ storage. To calculate the downward progression of the fingers, we consider the distance from the gas-water interface to the tip of the longest finger for any time, presented in **Fig. 20a)**. The vertical depth of investigation is quite similar for the CF, $CaCl_2$ 1M, and $MgCl_2$ 1M solutions. However, at higher concentrations ($CaCl_2$ 2M solution and $MgCl_2$ 2M solution), we observe a significant decrease in the vertical progression of fingers. Using NaCl with varying concentrations, Thomas et al. observed a similar effect [38]. Increasing salinity slows down the convective flow, causing less $CO_2$ to dissolve, consequently reducing the vertical depth of investigation. It should also be noted that dipping can reduce the vertical depth of investigation, as it promotes lateral mixing over vertical travel.

Another interesting phenomenon can be observed while considering the earlier part of the experiments, indicated by the red squared box in **Fig. 20(a)**, also magnified as presented in **Fig. 20(b)**. As shown in **Fig. 20(b)**, although the higher salt concentration in solution leads to a less vertical progression of fingers, the temporal finger evolution is higher at the earlier stage, also clearly observable from the velocity of finger evolution shown in **Fig. 21**. This indicates that the presence of salt enhances the diffusive flux which is the dominant mass-transfer method at an earlier stage. However, as time progresses, convection flows become the predominant flow causing the finger progression rate to slow down. This experimental investigation is also in perfect agreement with Kim and Kim's observation from the numerical simulation [41].

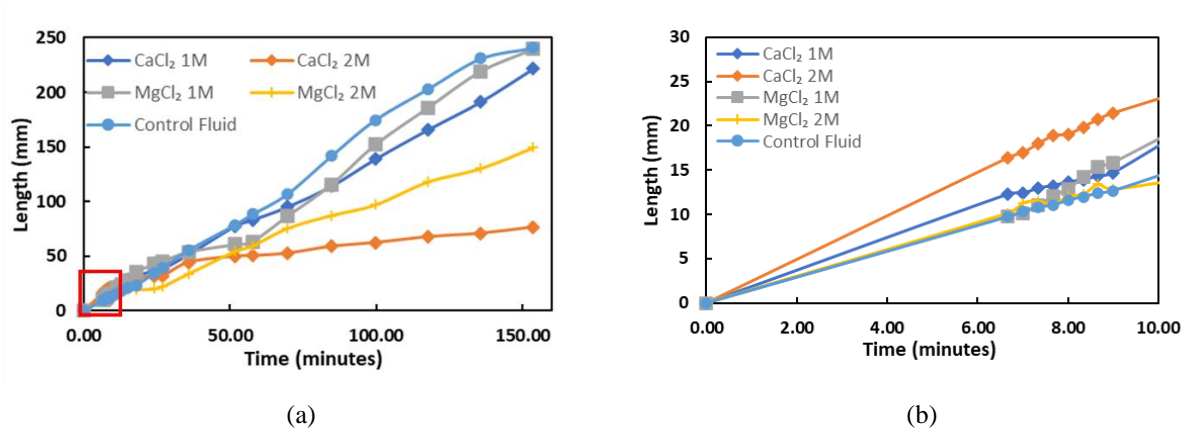

**Figure 20.** (a) Temporal evolution of the vertical progression of fingers for different cases (b) Magnified view of the red squared region

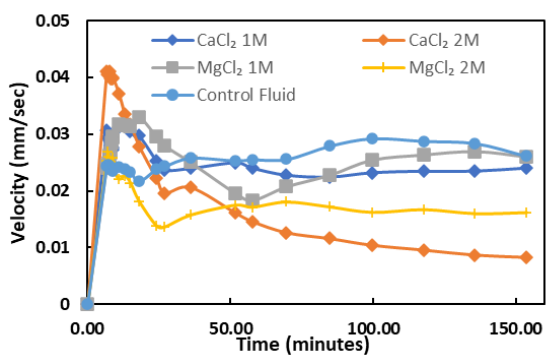

**Figure 21.** The velocity of vertical finger evolution for different cases

**Fig. 22** presents the result of image analysis for the salts with different concentrations, presented in terms of mean finger length, the number of fingers formed, mean wavelength, and standard deviation of finger length. We observe a slow increase in the mean finger length at the beginning of the experiment, and the number of fingers goes down as the neighboring fingers merge. However, after around 36 minutes, the nascent fingers with short finger lengths start to appear at the gas-water interface, which reduces the mean finger length. Moreover, the mature fingers start to merge, resulting in fewer dominant fingers, further decreasing the mean finger length, which is the weighted average of all the fingers. However, it should be noted that despite our observation of a reduction in the mean finger length, there is still vertical progression of fingers throughout the experiment, as evident in **Fig. 20**. This trend continues throughout the latter part of the experiment; as the larger finger merges and forms the oval shape formation, there is a noticeable decrease in the mean finger length, also causing a sharp increase in the standard deviation. Moreover, as the contributing factor in determining mean finger length is the vertical finger travel, the lateral mixing of fingers due to the dipping angle also causes a reduction in the mean finger length.

The convective finger formation and dissolution rate depend significantly on the type and concentration of salt in the CF. **Fig 22b** shows that the number of fingers formed is considerably slowed in case of a higher concentration of $CaCl_2$. There is a sharp increase in the number of fingers formed between 52 and 58 minutes since the fingers on the right side became apparent in that period. This shows that the $CO_2$ dissolution rate on the right side is significantly lower than on the left. Both the injection point and the presence of a dipping angle can cause this variance in dissolution rate.

**Fig. 22** shows that the number of fingers formed for MgCl$_2$ is almost similar to that of CaCl$_2$. The general trend of the parameters considered for image analysis looks identical for the same concentration of CaCl$_2$ and MgCl$_2$, especially in the earlier part of the experiment. In the latter half of the experiment, we observed the fingers in MgCl$_2$ having a higher mean finger length, particularly due to the faster vertical travel of the finger on the left side of the cell. Moreover, since the finger formation in the right side of the cell is significantly slow in all the cases, we observe a higher standard deviation in the later part of the experiment for the MgCl$_2$ experiments. Furthermore, the lower standard deviation of fingers in both the 2M solution cases can be attributed to the higher salinity and dipping angle preventing vertical travel and promoting lateral mixing of the fingers.

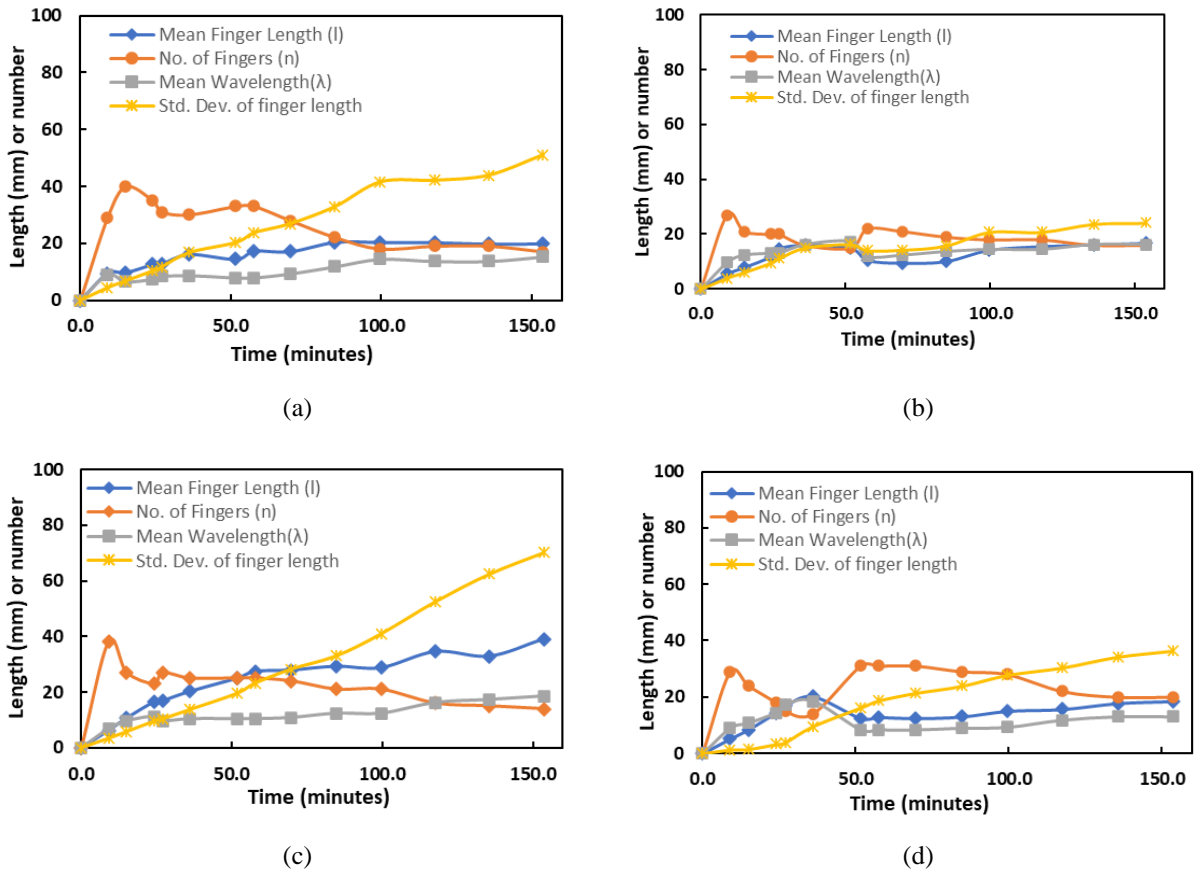

**Figure 22.** Number of fingers formed, wavelength and finger length parameters for (a) 1 mole CaCl$_2$ dissolved in Control Fluid – CO$_2$ introduced to the side of the cell and (b) 2 mole CaCl$_2$ dissolved in Control Fluid – CO$_2$ introduced to the side of the cell (c) 1 mole MgCl$_2$ dissolved in Control Fluid – CO$_2$ introduced to the side of the cell (d) 2 mole MgCl$_2$ dissolved in Control Fluid – CO$_2$ introduced to the side of the cell

## 5. Discussion

The structural morphology (formation dipping angle, pore-networks), geological properties (rock porosity, permeability, presence of fractures or barriers, etc.), and the accurate geochemical composition of the storage site play a significant role in transporting and storing supercritical $CO_2$ into the brine in dissolved form. Therefore, a fundamental understanding of these parameters and their role in convective $CO_2$ transport is essential to optimize the existing $CO_2$ storage operations and select appropriate new $CO_2$ storage sites. In addition, understanding the horizontal migration of supercritical $CO_2$ plumes due to structural dipping and point of injection is essential to reduce the chance of $CO_2$ leakage from fractured caprocks. The empirical observations of the parameters made in our study are summarized and presented in **Table 3**. This study critically evaluates these parameters to provide some important insights, which are discussed below:

## 5.1. Effect of salt concentration and dipping angle on convective dissolution

Our study showed that the presence of salts could significantly slow the convective flow, and the convective flow rate can depend on the salt type and its concentration. Although the convective flow is slowed down, we observed enhanced dissolution flux, the dominant mass-transfer method at the earlier stage of $CO_2$ storage. Furthermore, although the convective flow is slowed down, we observed an earlier onset time of convection in the presence of salt (**Table 2**). Earlier convection is important in $CO_2$ storage since it promotes faster $CO_2$ migration downwards in brine, leaving less risk of leakage. Moreover, we observed that different salts could cause a change in the convective flow pattern both in homogeneous and heterogeneous cases. Comparing the convective $CO_2$ flow pattern for without and with salts with different concentrations, we can conclude that lower salinity storage sites are more favorable for safe $CO_2$ storage, despite benefits like faster dissolution and earlier onset time of convection.

**Table 3.** Empirical observations on the parameters observed

| Parameters Observed | Empirical Observations |
|---|---|
| $CO_2$ Injection Point | • Injection point near boundary has different convective flow pattern than injection at the middle |
| Salinity | • Enhances the diffusive flux<br>• Slows down the convective flow<br>• Reduces $CO_2$ dissolution rate with increasing salinity |
| Dipping | • Promotes lateral mixing compare to vertical finger travel<br>• Reduces $CO_2$ storage with increasing dipping angle (however, the effect is not as significant as salinity) |
| Fractures | • Promotes high spatial $CO_2$ plume spreading<br>• Highly fractured regions have a high risk of $CO_2$ leakage during storage |
| Heterogeneity | • Different convection flow patterns are possible based on heterogeneous patterns<br>• Uneven $CO_2$ sweep with large unaffected regions |

Even though most of the experimental studies and simulations generalize geological storage sites to be horizontal, the majority of the sites have gradients as an effect of diagenesis and other geological phenomena. Although the effect of dip angle is less significant than that of salinity, the presence of the formation dip can significantly impact the space migration of $CO_2$ mixed fluid flow, as observed in our study. Due to the dominance of lateral mixing of $CO_2$ fingers over vertical travel, where dipping is involved, the depth of investigation is reduced for storage sites with a formation angle. A larger formation dip angle is not conducive to $CO_2$ geological storage as it promotes spatial $CO_2$ migration, risking long-term $CO_2$ storage. Further experimental studies on the effect of dipping angles on the depth of investigation can be done at a field-scale level for better theoretical benchmarking.

**5.2. Effect of flow barriers (boundary, heterogeneities, and fractures) on convective dissolution**

In our study, the injection point near the boundary had a considerably different convective flow pattern than the injection in the middle. From this visual investigation, we can infer that the vertical faults present at the storage site can cause a change in the convective flow pattern. Additionally, our characterization of convective flow for heterogeneities with fractures reveals how the $CO_2$ plume can migrate/spread through the fractures. Since the fractured regions can be highly conductive and increase $CO_2$ spatial spreading, fractures are not conducive to $CO_2$ storage. In addition, the formation of preferential channels for $CO_2$ acidified fluid flow is also observed. However, it should be noted that the channel growth and preferential fluid flow are more complex in storage sites and thus needs more attention.

Furthermore, the convective-flow pattern and the effective permeability of the storage site can vary significantly based on the fracture patterns. Therefore, the fraction patterns used to derive correlations will only apply to reservoirs with a similar pattern, showing the importance of considering all possible fracture network characteristics to critically estimate and model the complex fracture geometry for a particular storage site.

**5.3. Limitations of the study**

Reproducibility of the experiments needs careful control of experimental parameters, as any minor change can affect the convective flow evolution. Furthermore, it is worth noting that uneven aperture size due to the thickness distribution of the window glass sheets used in the Hele-Shaw cell can cause different permeable regions, causing unexpected convective flow. Thus, care must be taken to ensure a similar aperture throughout the cell for homogeneous cases.

**6. Conclusion**

This study provided a visual investigation into the effects of different parameters that dictates the $CO_2$ geologic storage: injection point and pressure, presence of salts with varying concentration, and presence of heterogeneities and dipping angle using the Hele-Shaw cell. Moreover, quantitative results were provided, which can be used for theoretical modeling. The visual investigation into the presence of heterogeneities and dipping angle shows how the variation in formation geometry dictates the flow of $CO_2$-acidified brine in the subsurface region. Our results show that the favorable conditions for $CO_2$ geological storage include low salinity, absence or low reservoir dipping, and a limited number of fractures. Furthermore, the convective-dissolution rate varies based on the presence of different salts; therefore, knowing the accurate chemical composition of the fluid in the storage site will result in accurate storage quantification. We observed the onset time of convection at around 135 s with a corresponding finger wavelength of 3.56 mm. In the presence of salt, the convection initiated at about 105 s with a corresponding finger wavelength of 6.33 mm. This difference shows that the initial diffusion is higher in brine with high salt concentrations but does not translate to enhanced convection and dissolution. The $CO_2$ dissolved area for different salts, indicated by the pH-depressed region, are considerably smaller than when no salt is present, and it follows the order of $CaCl_2$ 2M solution < $MgCl_2$ 2M solution < $CaCl_2$ 1M solution < $MgCl_2$ 2M solution. A factor not considered in this study but worthy of further investigation is the synergistic effect of multiple salts in saline aquifers which can be studied by creating solutions with a chemical composition similar to the brine by mixing several salts. By identifying the critical factors controlling the convective mixing, this study also provides new insights into the possible line of future work on the transport mechanisms during dissolution-driven $CO_2$ convective flow.

**Table of Abbreviations**

| | |
|---|---|
| 2D | Two Dimensional |
| 3D | Three Dimensional |
| BCG | Bromocresol Green |
| $CaCl_2$ | Calcium Chloride |
| CCS | Carbon Capture and Storage |
| CF | Control Fluid |
| $CO_2$ | Carbon Dioxide |
| $KMnO_4$ | Potassium Permanganate |
| LIF | Laser-Induced Fluorescence |
| NaCl | Sodium Chloride |
| PIV | Particle Image Velocimetry |
| PVT | Pressure-Volume-Temperature |
| Ra | Rayleigh number |
| RC | Rayleigh convection |
| RGB | Red-Green-Blue |
| UV | Ultraviolet |